# ProvSQL: A General System for Keeping Track of the Provenance and Probability of Data

Aryak Sen*, Silviu Maniu*†, Pierre Senellart‡†
*Univ. Grenoble Alpes, CNRS, Grenoble INP, LIG Grenoble, France
†CNRS@CREATE LTD, Singapore
‡DI ENS, ENS, CNRS, PSL University & Inria, Paris, France

***Abstract*—We present the data model, design choices, and performance of ProvSQL, a general and easy-to-deploy provenance tracking and probabilistic database system implemented as a PostgreSQL extension. ProvSQL's data and query models closely reflect that of a large core of SQL, including multiset semantics, the full relational algebra, and aggregation. A key part of its implementation relies on generic provenance circuits stored in memory-mapped files. We propose benchmarks to measure the overhead of provenance and probabilistic evaluation and demonstrate its scalability and competitiveness with respect to other state-of-the-art systems.**

## I. Introduction

As data is increasingly used for decision making and training machine learning models, it is crucial to account for two of its main inherent challenges. The first is that *sources matter*. Indeed, more and more importance is given to being able to trace final results back to the original data, to ensure that the results being presented to final users represent the truth. Secondly, in many cases data itself is *uncertain*, be it from the data collection process or as a result of data processing, such as using machine learning. To tackle these issues, it is important to have systems able to keep track of the *provenance* of data, and are also capable to *reason* about its uncertainty.

Data provenance – tracking data throughout a transformation process – has a long history in research, especially within relational data management. The main aims of data provenance are to keep track of *where* results come from, and to show *why* and *how* these results are computed. Early works on provenance [1], [2] reflected on the importance of determining the origin of query output, specifically in the form of *where* and *why* provenance. A related concept is that of data lineage [3], introduced in the Trio system for uncertain data management. A breakthrough was achieved through the *provenance semiring* framework [4] which concisely models many different forms of provenance, including why-provenance and Trio's lineage. In this framework, relations are *annotated* with elements of semirings and relational algebra operators are mapped to semiring operations. Extensions of this framework to more complex queries [5]–[7] have since been proposed.

Managing data uncertainty was another parallel line of research. Uncertainty was initially accounted for via concepts such as NULL values [8] and c-tables [9]. More recently, tuple-independent databases, where each tuple is independently annotated with a probability value, were introduced by Dalvi and Suciu [10], and extended to block-independent databases in [11], as a simple probabilistic model of uncertainty. Even in these simple frameworks, *probabilistic query evaluation*, i.e., computing the probability of a query result, is a #P-hard problem. Some tractable cases are however known: when queries have some specific properties (e.g., *safe queries* over tuple-independent databases [12]), or when data itself has a tree-like structure [13], captured by the notion of treewidth [14]. We are focusing in this work on the probabilistic database approach to modeling uncertainty, though we note that other approaches exist in the literature, e.g., [15], [16].

Though most of the works cited so far are theoretical in nature, various systems implement some form of provenance and probability evaluation or both: examples include Trio [3], Orchestra [17], Orion [18], MayBMS [19], Perm [20], and more recently GProM [21]. Except GProM, these systems are unfortunately unmaintained, have become hard to deploy or are even defunct. None of these systems provide a generic tool able to both keep track of the probability and provenance of data, for a wide variety of provenance frameworks.

In this article, we explain how the rich literature on provenance and probabilistic query evaluation is a basis for the implementation of ProvSQL, a plugin for the PostgreSQL database management system. ProvSQL aims to provide a generic, easy-to-deploy, and scalable solution to store and evaluate data provenance and probabilities. A proof-of-concept of ProvSQL was the subject of an early demonstration paper [22] in 2018; significant changes were brought to ProvSQL since: a complete architecture change (described in Section V-A), support for aggregation queries, computation of probability relying on tree decomposition techniques (see Section V-D), etc. ProvSQL was used as a provenance computation tool in several lines of research, by a variety of authors [23]–[28].

This paper is the first to provide a comprehensive presentation of its data model, query evaluation approach, and implementation aspects, as well as its real-world performance. Specifically, we provide the following contributions:

(i) We detail how the **theoretical results on provenance and probability are applied to real-world systems** using the SQL data model, using a generic representation of provenance and relying on a multiset semantics, and by evaluating the probability of Boolean provenance formulas for probabilistic query evaluation. We also discuss practical implementations of extensions to the basic semiring model via the monus operator [5] and semimodules for aggregates [6].

(ii) We present **design choices in ProvSQL** for provenance and probability computation, such as rewriting queries for provenance tracking, memory-mapped storage of provenance circuits, and knowledge compilation for probability computation.

(iii) To enable fair comparison with other systems and to quantitatively assess the performance of ProvSQL in real-world scenarios, we **propose publicly available benchmarks**, inspired from the TPC-H relational query benchmark, for both provenance and probabilistic query evaluation.

(iv) We **evaluate the performance of ProvSQL** on these benchmarks, both by itself (by measuring the provenance and probability overhead ProvSQL adds to PostgreSQL) and by comparing it to other systems that provide some of the features of ProvSQL. Specifically, we compare to GProM for provenance management and MayBMS for probabilistic query evaluation.

Our findings show that ProvSQL handles provenance and probabilistic query evaluation at scale (on multi-gigabyte databases), with reasonable overhead. Performance varies by query, leaving room for optimization. Compared with similar systems, ProvSQL supports a larger subset of SQL (including multiset semantics and aggregation) and works with arbitrary semirings. Its compact provenance circuits let it scale better than GProM. For probability computation, MayBMS is competitive with ProvSQL *on those queries MayBMS support*; however, ProvSQL often performs similarly even though it does not use any query-based optimization as MayBMS does.

ProvSQL is available as open source[1]. A companion repository[2] complements this paper, including benchmark scripts, additional details, proofs of all results, and links to formal definitions and proofs for the Lean proof assistant (also identified and linked to by L∃∀N throughout the paper).

## II. Related Systems

Trio [3] was an early system for managing uncertain data, focusing on the representation on different forms of uncertainty, including probabilities. It does not support computation of arbitrary marginals, which is at the core of modern probabilistic databases. Though it adopts a mediator approach, it is tied to specific and obsolete versions of PostgreSQL (8.2 or 8.3).

Obsolescence is a common problem for systems of this era: the provenance tracking system Perm [20] and the probabilistic datbase system Orion [18] are respectively unmaintained forks of PostgreSQL 8.3 and 8.4; the distributed Orchestra system [17], which provides an early implementation of provenance semirings, cannot be compiled because some of its dependencies are on servers that have disappeared; MystiQ [29], implementing safe query plan evaluation, requires Java 5.0, which reached end-of-service in 2009.

MayBMS [19][3] is a probabilistic database system implementing the general and compact model of U-relations [30]. Its query evaluator, SPROUT [31] was designed for efficiency and is in particular able to exploit the structure of *safe* queries [12]. It was developed as a fork of PostgreSQL 8.3, which is obsolete and hard to deploy. Some effort has been made in keeping the system compilable, however, though it requires using a virtual machine running an older operating system. It does not support provenance semirings. We use MayBMS in our experiments as a state-of-the-art system for probabilistic query evaluation.

GProM [21][4] is a middleware layer that adds provenance support to various database backends (in particular, Oracle and PostgreSQL). It translates declarative queries with provenance requirements into SQL code, which is subsequently executed by the backend database system. GProM supports the capture of provenance for SQL queries, as well as some Datalog queries. Some features not present in ProvSQL are support for transactions and PROV-JSON serialization [32]. On the other hand, in contrast with ProvSQL, it does not support probabilistic query evaluation, arbitrary semiring provenance, or provenance of non-monotone queries. As it is modern, actively maintained, and feature-rich, we use it as a comparison point in Section VI.

ProbLog [33][5] is an actively maintained and easy-to-use probabilistic programming system, inspired from Prolog and Datalog. SQL queries and probabilistic databases can be encoded into ProbLog in a straightforward way. Data can also be stored in a SQLite database and ProbLog uses knowledge compilation to compute probabilities of program outputs. However, our experiments showed it does not scale, as even though the data is stored in a database, it is loaded in main memory when needed by the query evaluator, and none of the usual query optimization infrastructure is used. Indeed, on the simplest query of our benchmark (query 1 of $\mathcal{Q}^{\text{cust}}$, see Section VI), ProbLog ran out-of-memory on the smallest database scale factor used in our experiments (1 GB). On a database 10 times smaller, it did not complete after 5 hours.

## III. Extended Relational Algebra

We now present the query language underlying ProvSQL; it is an extension of the relational algebra with multiset (or *bag*) semantics, allowing for aggregation. We first give basic definitions and notation about multisets.

**Definition 1.** *A* multiset *$m$ over a set $V$ is a function $V \to \mathbb{N}^*$. It is is* finite *if it has finite domain. For $v \in V$, we note $v \in m$ if $m(v) > 0$ and $v \notin m$ otherwise. Given two multisets $m_1$ and $m_2$ over $V$, we define the* multiset union *of $m_1$ and $m_2$ as $m_1 \uplus m_2 : x \mapsto m_1(x) + m_2(x)$ and the* cross product *of $m_1$ and $m_2$ as the multiset over $V^2$ defined by $m_1 \times m_2 : (x_1, x_2) \mapsto m_1(x_1) \times m_2(x_2)$. Any set $V$ can be seen as the multiset over $V$ where $m(v) = 1$ for all $v \in V$.*

*A finite multiset with domain $\{a_1, \dots, a_n\}$ can be written in the form $\left\{\!\!\left\{ \underbrace{a_1, \dots, a_1}_{m(a_1)\text{ times}}, \dots, \underbrace{a_n, \dots, a_n}_{m(a_n)\text{ times}} \right\}\!\!\right\}$. Finally, we also use the notation $\{\!\{ f(x) \mid x \in m \}\!\}$ or $\biguplus_{x \in m} \{\!\{ f(x) \}\!\}$ for a finite multiset $m$ with domain*

[1] https://github.com/PierreSenellart/provsql

[2] https://github.com/Aryak320/benchmark_suite

[3] https://maybms.sourceforge.net/

[4] https://github.com/IITDBGroup/gprom

[5] https://dtai.cs.kuleuven.be/problog/

$\{a_1, \ldots, a_n\}$ *and some function $f$ over $V$ to mean the multiset*

$$\left\{\!\!\left\{ \underbrace{f(a_1), \ldots, f(a_1)}_{m(a_1)\ \textit{times}}, \ldots, \underbrace{f(a_n), \ldots, f(a_n)}_{m(a_n)\ \textit{times}} \right\}\!\!\right\}.$$

We define an algebra for queries over relational databases under the multiset semantics which is as close as possible to a simple subset of SQL as dealt with by standard DBMSs. Our algebra is based on the relational algebra with set semantics typically used in database theory [34] but includes operators to explicitly control duplicate elimination as in [35], [36] and to express aggregation [37].

To simplify the presentation, we adopt an ordered, unnamed, and untyped perspective: attributes are referred to by their positions within the relation, do not have a name, and are assumed to be from a universal domain $\mathcal{V}$. In practical SQL implementations, attributes do have names and types, but it is straightforward to propagate them throughout query execution.

Let us fix a set $\mathcal{L}$ of relation labels and a set $\mathcal{V}$ of values. A database schema $\mathcal{D} : \mathcal{L} \to \mathbb{N}$ assigns arities to a *finite* set of relation labels. A *relation of arity $k \in \mathbb{N}$* is a *finite multiset* of tuples over $\mathcal{V}^k$. An instance of a database schema $\mathcal{D}$, or a database over $\mathcal{D}$ for short, is a function mapping each relation name $R$ in the domain of $\mathcal{D}$ to a relation of arity $\mathcal{D}(R)$.

A *positional index* is an expression of the form "#$i$" for $i \in \mathbb{N}^*$. A *term* is any expression involving positional indices and values from $\mathcal{V}$, combined using arbitrary operators and functions over values in $\mathcal{V}$. The *max-index* of a term is the maximum of all $i$ such that "#$i$" appears in the term (or 0 if none appear). Given a tuple $u = (u_1, \ldots, u_k)$ and a term $t$ of max-index $\leqslant k$, we define $t(u)$ as the value of $\mathcal{V}$ obtained by replacing in $t$ every positional index #$i$ with $u_i$ and evaluating the whole expression.

Given a database schema $\mathcal{D}$, we recursively define the language $\mathsf{RA}_k$ of relational algebra queries of arity $k \in \mathbb{N}$ as follows L∃∀N:

**relation** for any relation $R$ in the domain of $\mathcal{D}$, $R \in \mathsf{RA}_{\mathcal{D}(R)}$;

**projection** for $k \in \mathbb{N}$, $q \in \mathsf{RA}_k$, $t_1, \ldots, t_n$ terms of max-index $\leqslant k$, $\Pi_{t_1,\ldots,t_n}(q) \in \mathsf{RA}_n$;

**selection** for $k \in \mathbb{N}$, $q \in \mathsf{RA}_k$, and $\varphi$ a Boolean combination of (in)equality comparisons between terms of max-index $\leqslant k$, $\sigma_\varphi(q) \in \mathsf{RA}_k$;

**cross product** for $k_1, k_2 \in \mathbb{N}$, $q_1 \in \mathsf{RA}_{k_1}$, $q_2 \in \mathsf{RA}_{k_2}$, $q_1 \times q_2 \in \mathsf{RA}_{k_1+k_2}$;

**multiset sum** for $k \in \mathbb{N}$, $q_1, q_2 \in \mathsf{RA}_k$, $q_1 \uplus q_2 \in \mathsf{RA}_k$;

**duplicate elimination** for $k \in \mathbb{N}$, $q \in \mathsf{RA}_k$, $\varepsilon(q) \in \mathsf{RA}_k$;

**multiset difference** for $k \in \mathbb{N}$, $q_1, q_2 \in \mathsf{RA}_k$, $q_1 - q_2 \in \mathsf{RA}_k$;

**aggregation** for $k \in \mathbb{N}$, $q \in \mathsf{RA}_k$, distinct $(i_j)_{1 \leqslant j \leqslant k}$, terms $t_1, \ldots t_n$ of max-index $\leqslant k$, functions $f_1, \ldots, f_n$ from finite multisets of values to values, $\gamma_{i_1,\ldots,i_m}[t_1 : f_1, \ldots, t_n : f_n](q) \in \mathsf{RA}_{m+n}$.

Some other operators can be seen as syntactic sugar: **join** $q_1 \bowtie_\varphi q_2 \stackrel{\text{def}}{=} \sigma_\varphi(q_1 \times q_2)$; **set union** $q_1 \cup q_2 \stackrel{\text{def}}{=} \varepsilon(q_1 \uplus q_2)$.

The semantics $[\![\cdot]\!]^I$ of queries on an instance $I$ over $\mathcal{D}$ is defined as expected (see companion repository for details) for most operators. Note, however, that the definition of the multiset difference is not the usual one, and also not the one corresponding to the `EXCEPT ALL` operator of SQL, though it does correspond to the important `NOT IN` operator of SQL. This is mainly for practicality: the standard definition of multiset difference where $[\![q_1 - q_2]\!]^I(t) = \max(0, [\![q_1]\!]^I(t) - [\![q_2]\!]^I(t))$ leads to intractability of the provenance computation. The difference disappears in the set semantics: the semantics of $\varepsilon(q_1 - q_2)$ matches with that of `EXCEPT`.

Table I: The *Personnel* relation (derived from [38])

| id | name | position | city | |
|---|---|---|---|---|
| 1 | Juma | Director | Nairobi | $t_1$ |
| 2 | Paul | Janitor | Nairobi | $t_2$ |
| 3 | David | Analyst | Paris | $t_3$ |
| 4 | Ellen | Field agent | Beijing | $t_4$ |
| 5 | Aaheli | Double agent | Paris | $t_5$ |
| 6 | Nancy | HR | Paris | $t_6$ |
| 7 | Jing | Analyst | Beijing | $t_7$ |

Importantly, in this work, we assume that the aggregation operator, if it is used, is applied last (at the top-level operator), similarly to what is done in Section 3 of [6]; nested aggregation queries, discussed in Section 4 of [6], are implemented in ProvSQL, but as there is currently no clear semantics for semiring provenance for arbitrary semirings over nested aggregation queries, we leave them out-of-scope of this work.

**Example 2.** *Consider the instance $I$ of the relation Personnel given in Table I, containing names of people, their position and their city (remember attribute names are not part of the model, they are just given here for ease of reading). Ignoring for now the $t_i$'s, the relation has arity* 4. *Consider the following query:* "What are the cities where at least two persons are working?". *This can be expressed (without using aggregation) as:* $q_{\text{city}} \stackrel{\text{def}}{=} \varepsilon\left(\Pi_{\#4}\left(\textit{Personnel} \bowtie_{\#4=\#8\wedge\#1<\#5} \textit{Personnel}\right)\right)$. *Note the need for duplicate elimination $\varepsilon$ due to the multiset semantics. Then* $[\![q_{\text{city}}]\!]^I = \{\!\{\,(\text{Nairobi}), (\text{Paris}), (\text{Beijing})\,\}\!\}$.

## IV. Querying Annotated Relations

Now that our query language is fixed, we introduce the notion of annotated relations (Section IV-A) and the semantics of queries over such annotated relations in Section IV-B. This is based on annotated relations defined for provenance semirings as in [4], but with two key differences: we go beyond semirings for annotation, namely using the m-semirings from [5] and the $\delta$-semirings of [6], and, as in SQL, relations are multisets and not sets. We then show in Section IV-C how the semantics of provenance over annotated relations can be captured by query rewriting. Finally, we define probabilistic query evaluation and show how it can be performed in an intensional way relying on provenance (Section IV-D).

### A. Semirings and Beyond

Given some algebraic structure $\mathbb{K}$ for *annotations* (typically, a semiring or a generalization thereof) and some set $\mathcal{V}_\mathbb{K} \supseteq \mathcal{V}$ of *values* (typically, either $\mathcal{V}$ itself or an extension of $\mathcal{V}$ including $\mathbb{K}$-semimodules), a $\mathbb{K}$-*relation of arity $k \in \mathbb{N}$* over $\mathcal{V}_\mathbb{K}$ is a *multiset* of tuples over $(\mathcal{V}_{\mathbb{K}})^k \times \mathbb{K}$, often written $(u, \alpha)$ with $u \in (\mathcal{V}_{\mathbb{K}})^k$ and $\alpha \in \mathbb{K}$. A $\mathbb{K}$-instance of a database schema $\mathcal{D}$, or $\mathbb{K}$-database over $\mathcal{D}$ for short, is a function mapping each relation name $R$ in the domain of $\mathcal{D}$ to a $\mathbb{K}$-relation of arity $\mathcal{D}(R)$. [L∃∀N] Such $\mathbb{K}$-instances are typically denoted $\hat{I}$, and when we use such a notation $I$ then means the relation projected on the first $k$ columns, without the annotation.

We define in particular:

**Definition 3.** *A* semiring $(\mathbb{K}, \oplus, \otimes, \mathbb{0}, \mathbb{1})$ *is a set* $\mathbb{K}$ *of elements along with two binary operators over* $\mathbb{K}$ *(*$\oplus$ *and* $\otimes$*) and two distinguished elements of* $\mathbb{K}$ *(*$\mathbb{0}$ *and* $\mathbb{1}$*), such that:*

*(i)* $(\mathbb{K}, \oplus, \mathbb{0})$ *is a commutative monoid;*
*(ii)* $(\mathbb{K}, \otimes, \mathbb{1})$ *is a (non-necessarily commutative) monoid;*
*(iii)* $\otimes$ *distributes over* $\oplus$*:* $\forall a, b, c \in \mathbb{K}$, $a \otimes (b \oplus c) = (a \otimes b) \oplus (a \otimes c)$ *and* $(a \oplus b) \otimes c = (a \otimes c) \oplus (b \otimes c)$*;*
*(iv)* $\mathbb{0}$ *is annihilator for* $\otimes$*:* $\forall a\ \mathbb{0} \otimes a = a \otimes \mathbb{0} = \mathbb{0}$.

**Example 4.** *The following structures are semirings:*

**Counting semiring** $(\mathbb{N}, +, \times, 0, 1)$

**Boolean function semiring** *for a finite set* $X$, $(\mathcal{B}[X], \hat{\vee}, \hat{\wedge}, \hat{\bot}, \hat{\top})$ *where* $\mathcal{B}[X]$ *is the set of functions mapping valuations over* $X$ *(i.e., functions from* $X$ *to* $\{\bot, \top\}$*) to either* $\bot$ *(false) or* $\top$ *(true),* $f \hat{\vee} g$ *(resp.,* $f \hat{\wedge} g$*) is the function that maps a valuation* $\nu$ *to* $f(\nu) \vee g(\nu)$ *(resp.,* $f(\nu) \wedge g(\nu)$*), and* $\hat{\bot}$ *(resp.,* $\hat{\top}$*) is the constant function returning always* $\bot$ *(resp.,* $\top$*)*

**Why-provenance [2]** *for a finite set* $X$, $(2^{2^X}, \varnothing, \{emptyset\}, \cup, \Cup)$ *where* $\Cup$ *is defined by* $A \Cup B \stackrel{\text{def}}{=} \{a \cup b \mid a \in A, b \in B\}$

**Definition 5.** [L∃∀N] *A* semiring with monus $\mathbb{K}$*, or* m-semiring *is a semiring along with an extra binary operation* $\ominus$ *such that for all* $a, b, c \in \mathbb{K}$*: (i)* $a \oplus (b \ominus a) = b \oplus (a \ominus b)$*; (ii)* $(a \ominus b) \ominus c = a \ominus (b \oplus c)$*; (iii)* $a \ominus a = \mathbb{0} \ominus a = \mathbb{0}$.

The semirings from Example 4 can be extended with a monus operator to form an m-semiring: this was noted in [5] for $\mathbb{N}$ (where $\ominus$ is the truncated difference ($a \ominus b \stackrel{\text{def}}{=} \max(a-b, 0)$) and in $\mathcal{B}[X]$ (where it is $(a, b) \mapsto a \wedge \neg b$). For Why-provenance:

**Proposition 6.** [L∃∀N] *For a set* $X$, $(2^{2^X}, \varnothing, \{\varnothing\}, \cup, \Cup, \setminus)$ *is an m-semiring.*

For computing the provenance of aggregate queries, [6] introduces an extra operator: a *δ-semiring* is a semiring along with a unary operation $\delta$ such that: (i) $\delta(\mathbb{0}) = \mathbb{0}$; (ii) $\delta(\mathbb{1} \oplus \cdots \oplus \mathbb{1}) = \mathbb{1}$ whatever the number of $\mathbb{1}$s as input to $\delta$. For a discussion on choosing such a $\delta$ function, see [6]. A simple choice we will use for all semirings is the function that maps $\mathbb{0}$ to $\mathbb{0}$ and anything else to $\mathbb{1}$.

Finally, also to capture the provenance of aggregate queries, we will need to restrict which aggregate functions can be used:

**Definition 7.** *A* monoid aggregate function $f$ *over values in a set* $V$ *is a monoid homomorphism from the monoid of finite multisets of values of* $V$ *(with multiset union as monoid operation) to some monoid over* $V$*. In other words,* $f : (V \to \mathbb{N}^*) \to V$ *is a monoid aggregate function if there exists a monoid* $(V, \cdot, e)$ *such that* $f(\{\!\{\ \}\!\}) = e$ *and* $f(S \uplus T) = f(S) \cdot f(T)$.

**Example 8.** *The function* count*, that turns a multiset* $m$ *into the sum of all* $m(v)$ *for* $v$ *in the domain of* $m$ *is a monoid aggregate function from multisets of arbitrary values to* $(\mathbb{N}, +)$*. Similarly,* sum *is a monoid aggregate function from multisets over, say,* $\mathbb{Q}$ *to* $(\mathbb{Q}, +)$*, and* min *from multisets over* $\mathbb{Q}$ *to* $(\mathbb{Q}, \min)$.

### B. Algebra over Annotated Relations

We define the semantics $\langle\!\langle \cdot \rangle\!\rangle^{\hat{I}}$ of extended relational algebra queries on a $\mathbb{K}$-instance $\hat{I}$ over $\mathcal{D}$ by induction; for the definition to be meaningful, $\mathbb{K}$ needs to be a semiring; a m-semiring for the multiset difference operator; and a $\delta$-semiring for the aggregation operator. (We say that $\mathbb{K}$ is *appropriate* for $q$ if this is the case.) [L∃∀N]

**relation** for any relation label $R$ in the domain of $\mathcal{D}$, $\langle\!\langle R \rangle\!\rangle^{\hat{I}} \stackrel{\text{def}}{=} \hat{I}(R)$;

**projection** for $k \in \mathbb{N}$, $q \in \mathsf{RA}_k$, $t_1, \ldots, t_n$ terms of max-index $\leqslant k$, $\langle\!\langle \Pi_{t_1,\ldots,t_n}(q) \rangle\!\rangle^{\hat{I}} \stackrel{\text{def}}{=} \{\!\{ (t_1(u), \ldots, t_n(u), \alpha) \mid (u, \alpha) \in \langle\!\langle q \rangle\!\rangle^{\hat{I}} \}\!\}$;

**selection** for $k \in \mathbb{N}$, $q \in \mathsf{RA}_k$, and $\varphi$ a Boolean combination of (in)equality comparisons between terms of max-index $\leqslant k$, $\langle\!\langle \sigma_\varphi(q) \rangle\!\rangle^{\hat{I}} \stackrel{\text{def}}{=} \{\!\{ (u, \alpha) \mid (u, \alpha) \in \langle\!\langle q \rangle\!\rangle^{\hat{I}}, \varphi(u) \}\!\}$;

**cross product** for $k_1, k_2 \in \mathbb{N}$, $q_1 \in \mathsf{RA}_{k_1}$, $q_2 \in \mathsf{RA}_{k_2}$, $\langle\!\langle q_1 \times q_2 \rangle\!\rangle^{\hat{I}} \stackrel{\text{def}}{=} \{\!\{ (u, v, \alpha_1 \otimes \alpha_2) \mid (u, \alpha_1, v, \alpha_2) \in \langle\!\langle q_1 \rangle\!\rangle^{\hat{I}} \times \langle\!\langle q_2 \rangle\!\rangle^{\hat{I}} \}\!\}$;

**multiset sum** for $k \in \mathbb{N}$, $q_1, q_2 \in \mathsf{RA}_k$, $\langle\!\langle q_1 \uplus q_2 \rangle\!\rangle^{\hat{I}} \stackrel{\text{def}}{=} \langle\!\langle q_1 \rangle\!\rangle^{\hat{I}} \uplus \langle\!\langle q_2 \rangle\!\rangle^{\hat{I}}$;

**duplicate elimination** for $k \in \mathbb{N}$, $q \in \mathsf{RA}_k$, $\langle\!\langle \varepsilon(q) \rangle\!\rangle^{\hat{I}} \stackrel{\text{def}}{=} \bigcup_{u \mid \exists \alpha (u,\alpha) \in \langle\!\langle q \rangle\!\rangle^{\hat{I}}} \left\{ \left(u, \bigoplus_{\alpha \mid (u,\alpha) \in \langle\!\langle q \rangle\!\rangle^{\hat{I}}} \alpha \right) \right\}$;

**multiset difference** for $k \in \mathbb{N}$, $q_1, q_2 \in \mathsf{RA}_k$, $\langle\!\langle q_1 - q_2 \rangle\!\rangle^{\hat{I}} \stackrel{\text{def}}{=} \left\{\!\!\left\{ \left(u, \alpha \ominus \bigoplus_{\beta \mid (u,\beta) \in \langle\!\langle q_2 \rangle\!\rangle^{\hat{I}}} \beta \right) \middle| (u, \alpha) \in \langle\!\langle q_1 \rangle\!\rangle^{\hat{I}} \right\}\!\!\right\}$;

**aggregation** for $k \in \mathbb{N}$, $q \in \mathsf{RA}_k$, distinct $(i_j)_{1 \leqslant j \leqslant k}$, terms $t_1, \ldots t_n$ of max-index $\leqslant k$, monoid aggregate functions $f_1, \ldots, f_n$,

$$\langle\!\langle \gamma_{i_1,\ldots,i_m}[t_1 : f_1, \ldots, t_n : f_n](q) \rangle\!\rangle^{\hat{I}} \stackrel{\text{def}}{=} \Big\{ (v_1,\ldots,v_m, \hat{f}_1\big( \{\!\{ t_1(u) * \alpha \mid (u,\alpha) \in \langle\!\langle q \rangle\!\rangle^{\hat{I}}, (u_{i_1},\ldots,u_{i_m}) = (v_1,\ldots,v_m) \}\!\} \big), \ldots, \hat{f}_n\big( \{\!\{ t_n(u) * \alpha \mid (u,\alpha) \in \langle\!\langle q \rangle\!\rangle^{\hat{I}}, (u_{i_1},\ldots,u_{i_m}) = (v_1,\ldots,v_m) \}\!\} \big), \delta(\beta)) \Big| (v_1,\ldots,v_m,\beta) \in \langle\!\langle \varepsilon(\Pi_{\#i_1,\ldots,\#i_m}(q)) \rangle\!\rangle^{\hat{I}} \Big\}$$

where:

- "$*$" denotes a tensor product used to combine the data values of $\mathcal{V}$ with the $\delta$-semiring annotations from $\mathbb{K}$ in a semimodule $\mathcal{V}_{\mathbb{K}}$;
- for any $f_i$, $\hat{f}_i$ is a new aggregate function lifted from values in $\mathcal{V}$ to values in $\mathcal{V}_{\mathbb{K}}$

See [6] for details about provenance semimodules.

**Example 9.** *We return to Example 2 and to the instance of the* Personnel *table of Table I. The* $t_i$*'s are now interpreted as tuple annotations in some semiring* $\mathbb{K}$*, resulting in a* $\mathbb{K}$*-relation* $\hat{I}$*. Following the semantics of* $q_{\text{city}}$ *over annotated relations, we compute* $\langle\!\langle q_{\text{city}} \rangle\!\rangle^{\hat{I}}$ *as:*

| Nairobi | $t_1 \otimes t_2$ |
|---|---|
| Paris | $(t_3 \otimes t_5) \oplus (t_5 \otimes t_6) \oplus (t_3 \otimes t_6)$ |
| Beijing | $t_4 \otimes t_7$ |

*If we choose for $\mathbb{K}$ the semiring $\mathcal{B}[X]$ where $X = \{ t_1, \dots, t_7 \}$, the annotation for* Paris *is the Boolean function given by the formula $(t_3 \wedge t_5) \vee (t_5 \wedge t_6) \vee (t_3 \wedge t_6)$. This is a form of* Boolean provenance *[38]:* Paris *is in the result iff either both tuples representing* David *and* Aaheli*, or* Aaheli *and* Nancy*, or* David *and* Nancy*, are present.*

### C. Query Rewriting

The key way ProvSQL implements the semantics of the extended algebra over annotated relations is by *rewriting the query* over annotated relations to include in the query the necessary operations on provenance operations. The query can then be evaluated using the standard query evaluator of PostgreSQL. We employ the following 5 rewriting rules, which are applied inductively on an extended relational algebra formula: L∃∀N

(R1) for $k \in \mathbb{N}$, $q \in \mathsf{RA}_k$, $t_1, \dots, t_n$ terms of max-index $\leqslant k$, $\Pi_{t_1,\dots,t_n}(q)$ is rewitten to $\Pi_{t_1,\dots,t_n,\#(k+1)}(q)$

(R2) For $k_1, k_2 \in \mathbb{N}$, $q_1 \in \mathsf{RA}_{k_1}$, $q_2 \in \mathsf{RA}_{k_2}$, $q_1 \times q_2$ is rewritten to: $\Pi_{\#1,\dots,\#k_1,\#(k_1+2),\dots,\#(k_1+k_2+1),\#(k_1+1)\otimes\#(k_1+k_2+2)}(q_1 \times q_2)$.

(R3) For $k \in \mathbb{N}$, $q \in \mathsf{RA}_k$, $\varepsilon(q)$ is rewritten to: $\gamma_{1,\dots,k}[\#(k+1) : \bigoplus](q)$.

(R4) For $k \in \mathbb{N}$, $q_1, q_2 \in \mathsf{RA}_k$, $q_1 - q_2$ is rewritten to:

$$\Pi_{\#1,\dots,\#(k+1)}(q_1 \bowtie_{\#1=\#(k+2)\wedge\cdots\wedge\#k=\#(2k+1)} \varepsilon(\Pi_{\#1,\dots,\#k}(q_1) - \Pi_{\#1,\dots,\#k}(q_2)))$$
$$\uplus\ \Pi_{\#1,\dots,\#k,\#(k+1)\ominus\#(2k+2)}(q_1 \bowtie_{\#1=\#(k+2)\wedge\cdots\wedge\#k=\#(2k+1)} \gamma_{\#1,\dots,\#k}[\#(k+1):\bigoplus](q_2))$$

(R5) For $k \in \mathbb{N}$, $q \in \mathsf{RA}_k$, distinct $(i_j)_{1\leqslant j\leqslant k}$, terms $t_1, \dots, t_n$ of max-index $\leqslant k$, monoid aggregate functions $f_1, \dots, f_n$, $\gamma_{i_1,\dots,i_m}[t_1 : f_1, \dots, t_n : f_n](q)$ is rewritten to: $\gamma_{i_1,\dots,i_m}[t_1 * \#(k+1) : \hat{f}_1, \dots, t_n * \#(k+1) : \hat{f}_n, \#(k+1) : \delta(\bigoplus)](q)$.

Note that relation names, selection, and multiset sum operators are not rewritten. Using query rewriting to perform provenance evaluation has been proposed in previous works (see, in particular, Amsterdamer et al. [6], Fink and Olteanu [39] and Perm [20]). The two distinguishing aspects of this work is that we natively support the multiset semantics that SQL engines natively implement (requiring adaptation of rewriting rules for projection and explicit duplicate elimination) and that we support difference.

We show that these rewriting rules allow us to recover the exact semantics of the extended relational algebra over annotated relations:

**Theorem 10.** L∃∀N *Let $\mathcal{D}$ be a database schema, $q$ any extended relational algebra query over $\mathcal{D}$, $\mathbb{K}$ an appropriate algebraic structure, and $\hat{I}$ a $\mathbb{K}$-instance over $\mathcal{D}$. Let $\hat{q}$ be the query rewritten from $q$ by applying the rewriting rules (R1)–(R5) recursively bottom up. Then $\langle\!\langle q \rangle\!\rangle^{\hat{I}} = [\![\hat{q}]\!]^{\hat{I}}$.*

**Example 11.** *Returning to query $q_{\text{city}}$ from Example 2, let us trace the rewriting rules applied by ProvSQL bottom up (for space reasons, Personnel is abbreviated to P):*

$$
\begin{aligned}
q_{\text{city}} &= \varepsilon(\Pi_{\#4}(P \bowtie_{\#4=\#8\wedge\#1<\#5} P)) \\
&= \varepsilon(\Pi_{\#4}(\sigma_{\#4=\#8\wedge\#1<\#5}(P \times P))) \\
&\xrightarrow{(R2)} \varepsilon(\Pi_{\#4}(\sigma_{\#4=\#8\wedge\#1<\#5}(\Pi_{\#1,\dots,\#4,\#6,\dots\#9,\#5\otimes\#10}(P \times P)))) \\
&\xrightarrow{(R1)} \varepsilon(\Pi_{\#4,\#9}(\sigma_{\#4=\#8\wedge\#1<\#5}(\Pi_{\#1,\dots,\#4,\#6,\dots\#9,\#5\otimes\#10}(P \times P)))) \\
&\xrightarrow{(R3)} \gamma_1[\#2 : \bigoplus]( \\
&\qquad \Pi_{\#4,\#9}(\sigma_{\#4=\#8\wedge\#1<\#5}(\Pi_{\#1,\dots,\#4,\#6,\dots\#9,\#5\otimes\#10}(P \times P))))
\end{aligned}
$$

*If $\hat{q}_{\text{city}}$ is this rewritten query, one can check that $[\![\hat{q}_{\text{city}}]\!]^{\hat{I}} = \langle\!\langle q_{\text{city}} \rangle\!\rangle^{\hat{I}}$.*

### D. Probabilistic Query Evaluation

Annotated relations can be used for probabilistic query evaluation using the so-called intensional approach [40]. First, assume a finite set $X$ of variables, each variable $x \in X$ being assigned a probability $\Pr(x)$. Pr can be extended to a probability distribution over valuations over $X$, assuming independence of variables: $\Pr(\nu) = \prod_{x|\nu(x)=\top} \nu(x) \times \prod_{x|\nu(x)=\bot}(1 - \nu(x))$. This in turns extends to a probability distribution over Boolean functions: if $f \in \mathcal{B}[X]$ is a Boolean function over variables in $X$, then $\Pr(f) = \sum_{\nu|f(\nu)=\top} \Pr(\nu)$.

Consider a $\mathcal{B}[X]$-relation $\hat{I}$ of arity $k$. For any subinstance $\hat{J}$ of $\hat{I}$, the characteristic Boolean function of $\hat{J}$ within $\hat{I}$ is: $\Phi_{\hat{I}}(\hat{J}) \stackrel{\text{def}}{=} \bigwedge_{(u,\alpha)\in\hat{J}} \alpha \wedge \bigwedge_{(u,\alpha)\in\hat{I},(u,\alpha)\notin J} \neg\alpha$. The probability of $\hat{J}$ is then $\Pr(\Phi_{\hat{I}}(\hat{J}))$. For a query $q$ over $\hat{I}$ in $\mathsf{RA}_k$, we define the marginal probability that a tuple $t$ of arity $k$ appears in the output of $q$ as $\Pr(t \in q(\hat{I})) \stackrel{\text{def}}{=} \sum_{\hat{J}\subseteq\hat{I},t\in[\![q]\!]^J} \Pr(\hat{J})$.

Our semantics for extended relational algebra over $\mathbb{K}$-relations is compatible with probabilistic query evaluation:

**Theorem 12.** *For any finite set of variables $X$, probability distribution* Pr *over $X$, $\mathcal{B}[X]$-relation $\hat{I}$ and relational algebra query $q$ without aggregation, for any tuple $t$ with same arity as $q$, $\Pr(t \in q(\hat{I})) = \Pr\left(\bigvee_{(t,\alpha)\in\langle\!\langle q \rangle\!\rangle^{\hat{I}}} \alpha\right)$.*

The reason why this theorem is about queries without aggregation is that the result of aggregate queries are tuples whose values include annotations; for a similar result, we would need to talk about the distribution of data values, or some summary thereof (such as the expected value). This is out of scope of this work.

Combining Theorems 10 and 12, we obtain:

**Corollary 13.** *For any finite set of variables $X$, probability distribution* Pr *over $X$, $\mathcal{B}[X]$-relation $\hat{I}$ and relational algebra query $q$ without aggregation, for any tuple $t$ with same arity as $q$, if $\hat{q}$ is the query rewritten from $q$ by applying the rewritten rules (R1)–(R5) then $\Pr(t \in q(\hat{I})) = \Pr\left(\bigvee_{(t,\alpha)\in[\![\hat{q}]\!]^{\hat{I}}} \alpha\right)$.*

**Example 14.** *Returning again to Examples 2 and 9, assume we independently assign to each $t_i$ in our Personnel $\mathcal{B}[\{ t_1, \dots, t_7 \}]$-instance a probability $\Pr(t_i)$ with $\Pr(t_1) = 0.5$* *and* $\Pr(t_2) = 0.7$. *We can then compute the probability of* $q_{\text{city}}$ *results as, for instance for Nairobi:* $\Pr(t_1 \wedge t_2) = 0.5 \cdot 0.7 = 0.35$.

## V. Implementation in ProvSQL

Theorem 10 and Corollary 13 pave the way for a practical implementation of provenance management and probabilistic query evaluation in a SQL DBMS. Note that the data model follows as closely as possible the practical behavior of SQL engines, instead of the theoretical model of annotated relations used in most of the literature: multiset semantics, explicit duplicate elimination using **DISTINCT** or **GROUP BY**, ordered attributes within relations.

We now explain how ProvSQL is implemented. ProvSQL uses PostgreSQL's extension mechanism to change the behavior of the DBMS. This extension mechanism allows to provide user-defined functions (UDFs), which can be written either in PL/pgSQL, the application language of PostgreSQL, or in C, to implement provenance- and probability-related functionalities in C. The extension mechanism also allows to run *hooks* through a C interface at different key phases of query evaluation. ProvSQL uses a combination of SQL, PL/pgSQL, C, and C++ code interfacing with the C code, to implement its behavior.

### A. Annotated relations

As in Section IV-A, ProvSQL adds to every relation for which provenance needs to be tracked (which can be specified with a UDF) an extra *provsql* attribute to store the annotation. Instead of choosing a specific algebraic structure for the annotation ahead of time, this annotation is a generic *universally unique identifier (UUID)* [41]. We explain further on how these generic annotations can be specialized to specific (m)-semirings. Annotations of *base tuples* are randomly generated using the UUIDv4 standard and can be interpreted as abstract tuple identifiers; the 128-bit address space makes the collision probability vanishingly small. Annotations are built from these base annotations by using the (m-)semiring operations; to store resulting annotations in the same attribute, they are also UUIDs, this time computed following UUIDv5 by computing a SHA-1 hash formed from a fixed namespace, and a normalized description of the operator and its UUID operands. This guarantees, in particular, that annotations are generated in a deterministic manner: the same query over the same data will result in the exact same result, annotations included.

Retaining opaque UUIDs is not enough: one also needs to be able to efficiently determine how they relate to each other. This is done by storing a *provenance circuit* (a compact DAG representation of provenance expressions, first introduced in [42]) where each UUIDv4 points to a leaf gate representing inputs of the circuit and each UUIDv5 to an inner gate labeled by the operator and with children the operand UUIDs. Similarly, data values that are results of aggregate queries are represented as gates in the same circuit (with UUIDv5 representation) encoding their semimodule construction.

Storage of this provenance circuit in the DBMS is a challenge:

(i) It is an append-only data structure; except in cases where one wants to perform some clean-up, there is never the need of removing gates from the circuit, and a gate never needs to be updated.

(ii) It is a data structure which is updated at the time a query is evaluated, following the query rewriting approach of Section IV-C.

(iii) The content of this circuit needs to be stored in a persistent manner: UUIDs become uninterpretable without the circuit.

(iv) The circuit can become very large, as it contains one gate per tuple, and one gate per operation performed in a query.

(v) Access to the circuit needs to be as fast as possible.

Initial implementations of ProvSQL stored this circuit as an extra table within the database, which solved (iii)–(iv). But PostgreSQL tables are not optimized for append-only operations (i) and, most importantly, (ii) caused concurrency control issues. We then experimented with shared-memory storage of the circuit, which solved most issues but did not scale (iv) and made persistence (iii) an issue in case of failure.

Our final implementation provides a satisfactory solution to (i)–(v): The circuit is stored outside of the database, in *append-only files* that are properly indexed and *memory-mapped* so that the operating system can keep most often used fragments in RAM buffers. To avoid concurrency issues, access to these files is managed by a *single PostgreSQL worker process*, communicating with other backends through inter-process communication (with *pipes*). Finally, each backend process has a *small local cache* of most recently accessed circuit information.

### B. Query rewriting

The main way ProvSQL is keeping track of the provenance of data is by the query rewriting approach discussed in Section IV-C. To implement this within PostgreSQL, ProvSQL defines a custom hook at *query planning time* before the query is sent to PostgreSQL's planner. If the query involves annotated relations (i.e., tables with the special *provsql* attribute of UUID type), the query is analyzed to determine if it is part of the supported language (the extended relational algebra described in Section III, as represented in SQL). If not, an error message is displayed instead of performing inadequate provenance computation. If so, the query is rewritten as described in Section IV-C and then sent to the regular planner for continued processing and evaluation.

### C. Specialization to specific (m-)semirings

Annotations computed by ProvSQL are abstract UUIDs, with description in the circuit on how these are computed from base tuple UUIDs. One important point is that representation can be specialized to any arbitrary (m-)semiring and semimodule by applying homomorphisms: ProvSQL essentially works in the *universal* semiring (the how-semiring of integer polynomials, see [4]) and in the *universal* m-semiring (the free m-semiring, see [5]), from which there exist unique homomorphisms to any application (m-)semiring. Each algebraic structure of interest is compiled into a C++ class whose methods describe the different operations (⊕, ⊗, possibly ⊖ and $\delta$) of the algebraic structure. At a user's request, given a *provenance mapping* from base UUIDs to values in that algebraic structure (e.g., integers in the counting semiring), the relevant part of the circuit is retrieved from its memory mapped representation, and ProvSQL evaluates the circuit in a bottom-up manner to determine the actual semiring annotation of every tuple in the result of a query, using the methods provided by the class.

### D. Probability evaluation

Probabilistic query evaluation in ProvSQL relies on Corollary 13: provenance is computed in the same way as for provenance tracking, using query rewriting, and when a marginal probability computation is requested, we specialize the provenance to $\mathcal{B}[X]$ (given a mapping of base UUIDs to some Boolean function, e.g., one distinct variable per UUIDs to obtain a tuple-independent database) and then compute the probability of the corresponding Boolean circuit. Note that the computation of Boolean provenance, as it is crucial for probabilistic query evaluation, is further optimized from the regular retrieval of subcircuits from the the memory-mapped circuit.

ProvSQL implements different ways of computing the probability of a Boolean circuit (e.g., naïve enumeration of possible worlds, Monte-Carlo sampling, or some other approximation techniques such as WeightMC [43]). We focus here on its default computation method, which works in three steps:

(1) We determine whether the Boolean circuit is read-once [44], i.e., if every variable occurs only once; if so, its probability can be computed in linear-time.

(2) Otherwise, we use a heuristic algorithm [14] to quickly obtain a tree-decomposition of small width ($\leqslant 10$) of the circuit, if possible; if so, we use an algorithm from [45] to extract a deterministic and decomposable circuit from the Boolean circuit, on which the probability is computed in linear time.

(3) Otherwise, we encode the circuit in conjunctive normal form using Tseitin's transformation [46] and use an external knowledge compiler (by default d4 [47] but we also support c2d [48] and DSHARP [49]) as a black box to compile it into a deterministic and decomposable circuit, on which the probability is computed in linear time (but this new circuit may be exponential in size).

Recall that probabilistic query evaluation is intractable (#P-hard): the potential exponential blow-up of the last step is unavoidable.

### E. Other features

At the time of writing, ProvSQL also implements other features, which are outside of the scope of this paper: it supports provenance for non-terminal aggregate queries, following Section 4 of [6]; Shapley value computation and expected Shapley value computation as described in [50]; provenance of update operations [51] as proposed in [7]; expected value computation for **COUNT**, **MIN**, **MAX**, **SUM**, following the algorithms in [52]; and where-provenance [2], [22], which is a form of provenance not captured by semirings [53].

## VI. Experimental Evaluation

We start this section by presenting how we built our benchmark and on which SQL queries different systems are tested. We then evaluate the performance of ProvSQL for provenance and probability computation, and compare it to GProM [21] and MayBMS [19].

### A. Building the Benchmark

A natural candidate for benchmarking queries in database systems is the TPC benchmark suite[6]. We started with the TPC-H 3.0.1 database generator and its associated suite of queries. TPC-H is a standard benchmark for decision support systems, and includes 22 query templates that are typical of this kind of application, including some with high computational cost. The TPC-H schema is formed of 8 tables, 6 whose size increases proportionally to a provided *scale factor* (`lineitem`, `orders`, `part`, `supplier`, `customer`, and `partsupp`) and 2 which are fixed (`nation` and `region`). The scale factor allows to control the total size of the database. A scale factor of $k$ results in a database of roughly $k$ GB.

None of the systems tested currently support the functionalities present in all TPC-H queries. In particular none of them support provenance tracking or probability computations for subqueries in the **WHERE** clause or **ORDER BY** on the result of aggregate queries. As previously discussed, though recent versions of ProvSQL do support nested aggregation queries, in the absence of a clear semantics of provenance tracking for nested aggregation, we also consider such queries out-of-scope. Only 6 queries of TPC-H work as is in at least one tested system (as it happens, ProvSQL); this query set is denoted as $\mathcal{Q}^{\text{TPC}}$. This query set is unfortunately not suitable for GProM and MayBMS as all queries include aggregation operators, which these systems do not support.[7]

MayBMS and its query evaluator SPROUT [31] were also originally evaluated on modified TPC-H queries, and MayBMS includes a benchmark of 4 such queries, modified to remove unsupported operations[8]. We include them in a second part of our benchmark, denoted $\mathcal{Q}^{\text{TPC*}}$, along with TPC-H query 3, specifically modified to remove its aggregation operator (which GProM and MayBMS do not support).

$\mathcal{Q}^{\text{TPC}}$ and $\mathcal{Q}^{\text{TPC*}}$ are still not representative, in that they do not cover the full range of SQL features as implemented by provenance systems. We thus created an additional set $\mathcal{Q}^{\text{cust}}$ of 18 custom queries chosen to represent a wide variety of simple query patterns, but which cover important features of the SQL language. We relied on the DSQGEN query generator from the TPC-DS benchmark for generation because of its flexibility, but the queries are on the TPC-H schema as well. We used $\mathcal{Q}^{\text{cust}}$ and other subsets of our custom queries as query loads, executing them without restarting the Postgres server in between individual query runs, allowing caching. These 18 queries include different SQL features such as joins, varying number of aggregates, `EXCEPT`, `UNION`, `DISTINCT`, `GROUP BY`. Table II shows an overview of the features of all queries.

[6] https://www.tpc.org/information/benchmarks5.asp

[7] MayBMS supports limited forms of summaries for probabilistic query evaluation of aggregate queries, with the `ecount()` and `esum()` functions but not standard SQL aggregates. Similarly, GProm supports some form of aggregation [21], but not in a way compatible with semiring provenance.

[8] https://maybms.sourceforge.net/manual/index.html

Table II: Benchmark query features

| | $\mathcal{Q}^{\text{TPC}}$ | | | | | | $\mathcal{Q}^{\text{TPC*}}$ | | | | | $\mathcal{Q}^{\text{cust}}$ | | | | | | | | | | | | | | | | | |
|---|---|---|---|---|---|---|---|---|---|---|---|---|---|---|---|---|---|---|---|---|---|---|---|---|---|---|---|---|---|
| **Query** | 1 | 6 | 7 | 9 | 12 | 19 | 1 | 3 | 4 | 12 | 15 | 1 | 2 | 3 | 4 | 5 | 6 | 7 | 8 | 9 | 10 | 11 | 12 | 13 | 14 | 15 | 16 | 17 | 18 |
| # tables joined | 1 | 1 | 6 | 6 | 2 | 2 | 1 | 3 | 2 | 2 | 2 | 1 | 2 | 5 | 2 | 2 | 1 | 1 | 2 | 8 | 4 | 3 | 2 | 2 | 5 | 3 | 4 | 3 | 1 |
| # aggregates | 8 | 1 | 1 | 1 | 2 | 1 | 0 | 0 | 0 | 0 | 0 | 0 | 0 | 0 | 0 | 0 | 0 | 0 | 0 | 0 | 0 | 0 | 0 | 0 | 0 | 0 | 0 | 0 | 0 |
| UNION | ✗ | ✗ | ✗ | ✗ | ✗ | ✗ | ✗ | ✗ | ✗ | ✗ | ✗ | ✗ | ✗ | ✗ | ✗ | ✗ | ✗ | ✗ | ✗ | ✗ | ✗ | ✗ | ✗ | ✗ | ✗ | ✗ | ✗ | ✗ | ✓ |
| EXCEPT | ✗ | ✗ | ✗ | ✗ | ✗ | ✗ | ✗ | ✗ | ✗ | ✗ | ✗ | ✗ | ✗ | ✗ | ✗ | ✗ | ✗ | ✗ | ✗ | ✗ | ✗ | ✗ | ✗ | ✗ | ✗ | ✗ | ✓ | ✗ | ✗ |
| DISTINCT | ✗ | ✗ | ✗ | ✗ | ✗ | ✗ | ✗ | ✗ | ✗ | ✗ | ✗ | ✗ | ✓ | ✗ | ✗ | ✗ | ✗ | ✗ | ✗ | ✗ | ✗ | ✗ | ✗ | ✗ | ✗ | ✓ | ✗ | ✗ | ✗ |
| GROUP BY | ✓ | ✗ | ✓ | ✓ | ✓ | ✗ | ✓ | ✓ | ✓ | ✓ | ✓ | ✗ | ✗ | ✗ | ✓ | ✗ | ✓ | ✗ | ✗ | ✓ | ✓ | ✗ | ✗ | ✗ | ✓ | ✗ | ✗ | ✓ | ✗ |
| ORDER BY | ✓ | ✗ | ✓ | ✓ | ✓ | ✗ | ✗ | ✗ | ✗ | ✗ | ✗ | ✗ | ✗ | ✗ | ✗ | ✗ | ✗ | ✗ | ✗ | ✗ | ✗ | ✗ | ✗ | ✗ | ✗ | ✗ | ✗ | ✗ | ✗ |

Table III: System comparison: supported queries on the TPC-H 1GB dataset (Y – successful execution, O – out of memory, N – query not supported, T – timeout > 3000$s$)

| | $\mathcal{Q}^{\text{TPC}}$ | | | | | | $\mathcal{Q}^{\text{TPC*}}$ | | | | | $\mathcal{Q}^{\text{cust}}$ | | | | | | | | | | | | | | | | | |
|---|---|---|---|---|---|---|---|---|---|---|---|---|---|---|---|---|---|---|---|---|---|---|---|---|---|---|---|---|---|
| **Query** | 1 | 6 | 7 | 9 | 12 | 19 | 1 | 3 | 4 | 12 | 15 | 1 | 2 | 3 | 4 | 5 | 6 | 7 | 8 | 9 | 10 | 11 | 12 | 13 | 14 | 15 | 16 | 17 | 18 |
| ProvSQL (prov.) | Y | Y | Y | Y | Y | Y | Y | Y | Y | Y | Y | Y | Y | Y | Y | Y | Y | Y | Y | Y | Y | Y | Y | Y | Y | Y | Y | Y | Y |
| ProvSQL (prob.) | O | Y | O | Y | Y | Y | Y | Y | Y | Y | Y | Y | Y | Y | Y | Y | Y | Y | Y | O | Y | Y | Y | Y | Y | Y | Y | Y | Y |
| GProM | N | N | N | N | N | N | Y | Y | Y | Y | Y | O | Y | Y | Y | O | Y | Y | Y | Y | Y | Y | O | Y | Y | Y | N | Y | N |
| MayBMS | N | N | N | N | N | N | Y | Y | Y | Y | Y | Y | Y | Y | T | O | Y | Y | Y | O | Y | Y | Y | Y | Y | Y | N | Y | N |

Table III provides a summary of which queries are supported on every system for the lowest scale factor (1 GB): a "Y" means the query runs correctly, a "N" is an error due to unsupported operators, an "O" an out-of-memory error, and a "T" a timeout at 3 000 seconds. For ProvSQL, we distinguish between computation of provenance and probability evaluation. We observe that ProvSQL is able to process all three query sets; for probabilistic query evaluation, all but three run to completion, and none of those 3 failing queries work in MayBMS. Other than that, a number of queries supported by ProvSQL are either unsupported (aggregation, difference, union operators) or fail using MayBMS or GProM. GProM in particular is particularly sensitive to running out of memory.

## B. Experimental Setup

Benchmarks were executed on a Dell Precision 7560 workstation running Debian 12 with a 11th Gen Intel(R) Core(TM) i9-11950H processor and 64GB RAM. ProvSQL and GProM used PostgreSQL 16. As MayBMS cannot be deployed on modern systems, we had to run it under a virtual machine run on the same machine, provided with 50 GB RAM and 8 CPU cores; the VM ran Ubuntu 10.

We perform the benchmarks on the TPC-H 3.0.1 dataset with indexes on primary keys (unless otherwise mentioned) for scale factors 1 to 10, i.e., with databases of size 1 to 10 GB. For experiments involving probability computation, we modify the deterministic TPC-H database into a tuple-independent probabilistic database by assigning a probability of 0.5 to every tuple of the relations involved (note that algorithms used for exact probabilistic query evaluation in ProvSQL and MayBMS do not depend on the actual probability values of uncertain tuples, so another probability valuation would have given the same results in terms of running time).

We use `ANALYZE` to force PostgreSQL to collect statistics about tables once each database is fully loaded. Each experiment is run 10 times, with the database server restarted each time, to avoid fluctuations independent of the system's performance. Results presented are averaged over these 10 runs.

## C. Benchmarking ProvSQL

We start with evaluating the provenance tracking of ProvSQL on our benchmark queries, for three scenarios: (i) the overhead of adding provenance tracking, (ii) the scalability of different semiring instantiations, and (iii) the scalability of adding probability evaluation.

*Scalability and provenance overhead:* We compare the running time of queries over PostgreSQL with the running time under ProvSQL, with provenance tracking enabled, on TPC-H data of scale factors between 1 and 10. We are especially interested in the overhead ratio, computed as overhead = $\frac{t_{\text{provenance}}}{t_{\text{original}}}$. We run the experiment on the entire $\mathcal{Q}^{\text{cust}}$query set, and on individual queries from $\mathcal{Q}^{\text{TPC}}$. We show the results in Figure 1. Note that the blue curve corresponds to the overhead, on the right (linear) y-axis of each plot, while the other two curves use the left (logarithmic) y-axis.

On $\mathcal{Q}^{\text{cust}}$, Figure 1a shows that the overhead is near-constant with the dataset size, indicating that provenance tracking scales at a similar rate as regular query execution. This reflects the efficient implementation of provenance tracking by using provenance circuits in ProvSQL.

Results on $\mathcal{Q}^{\text{TPC}}$, Figure 1b, are of particular interest. Query 19 has a nearly constant overhead with little variation as the scale factor increases, characterizing efficient provenance tracking that scales independently of the database size. For queries 7 and 9, we observe an anomalous *decrease* in overheads at certain scale factors due to PostgreSQL's query optimizer adopting different execution plans when provenance tracking is

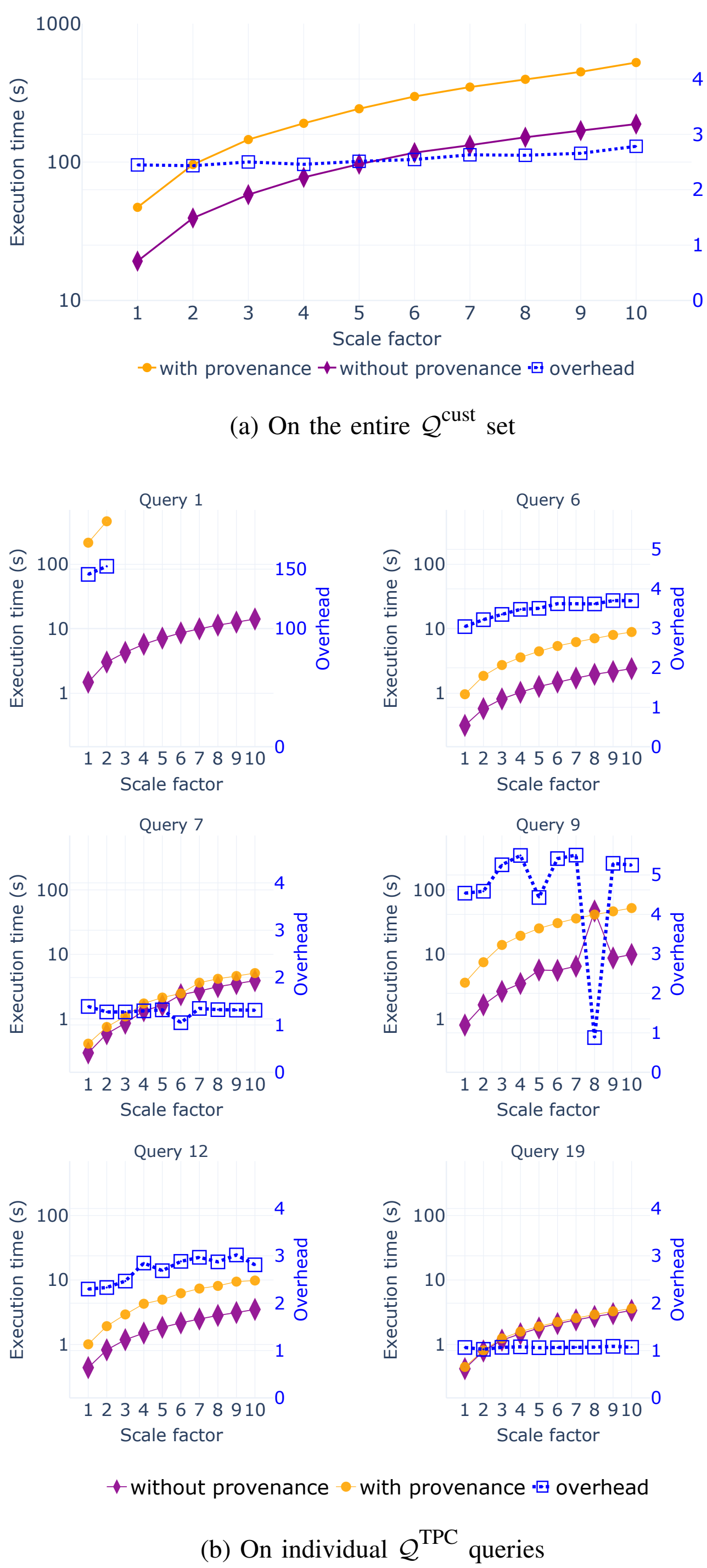

(a) On the entire $\mathcal{Q}^{\text{cust}}$ set

(b) On individual $\mathcal{Q}^{\text{TPC}}$ queries

Figure 1: Scalability and provenance overhead: execution times (left $y$-axis, log-scale) and overhead (right $y$-axis, linear scale), at various scale factors ($x$-axis)

enabled, which happen to be better for that particular query. For instance on query 9, the overhead at scale factor 8 completely disappears. Indeed, in certain cases, provenance computation can accidentally lead to performance improvement by guiding the PostgreSQL optimizer to adopt a faster execution strategy. The size of the TPC-H dataset incrementally increases with each scale factor and can cause changes in join selectivity, indexing effectiveness and the optimizer's cost estimates. For example, in queries 7 and 9 which have complex joins, provenance tracking may force the optimizer to adopt a more efficient join order. Except for these optimizer-driven fluctuations, queries 7 and 9 have a constant overhead. Queries 6 and 12 exhibit a high overhead that however remains under a factor of 4, indicating scalability challenges in comparison to other queries in this experiment. Query 6 does not involve any inner joins and operates exclusively on the largest table in the dataset, `lineitem`. The size of `lineitem` scales with the TPC-H scale factor and also contains attributes with skewed, uniform, and categorical distributions, making it a bottleneck for database optimizers in situations where provenance tracking involves queries with a high amount of filtering. For query 12, the high overhead may be because of the fact that it involves filtering conditions on aggregated tuples. We note that ProvSQL cannot scale beyond scale factor 2 for query 1 due to memory limitations; at scale factor 1, the number of gates created in the provenance circuit by ProvSQL for this query is 45.71M gates. For comparison, for query 7 it is 18.27k gates and for query 9 it is 0.99M gates. Query 1 is a relatively broad query having a single filter on dates filter along with 8 aggregations on the `lineitem` table, grouped on low-cardinality columns such as *l_linestatus* and *l_returnflag*, thus limiting scalability.

This experiment provides us with an important insight: while computing provenance annotations generally adds overhead, this overhead is limited and even constant in most cases. It also indicates avenues for optimization of provenance computation, especially by reducing intermediate data structures.

*Semiring provenance evaluation:* We evaluate in Figure 2 the overhead induced by evaluating the resulting provenance circuits on different semirings supported by ProvSQL (*counting*, *why-provenance*, as well as a pseudo-semiring that just serializes in a *formula*-based representation the provenance circuit), on the TPC-H database across scale factors 1 to 10. On the queries in $\mathcal{Q}^{\text{cust}}$ (Figure 2a) we observe linear trends, with the semiring instantiations across TPC-H scale factors 1 to 10. All three of the semirings introduce virtually the same overhead to provenance computation. The overheads observed are constant with the database size, as none of the queries in $\mathcal{Q}^{\text{cust}}$ have aggregates. This increased efficiency compared to aggregate provenance is due to the fact provenance computation for aggregates involves poly-sized overheads caused by the $K-$semimodule operations.

Figure 2b shows the overheads incurred for individual $\mathcal{Q}^{\text{TPC}}$ queries. We omit query 1 from $\mathcal{Q}^{\text{TPC}}$ in this experiment, as we just saw ProvSQL does not scale for query 1 even when we are only computing the provenance circuit. Except for query 9 the execution times for the semiring instantiations indicate minimal poly-sized overheads over circuit computation times.

*Comparing scalability and performance queries 7 and 9 of $\mathcal{Q}^{TPC}$:* We detail the functioning of queries 7 and 9 for different implementation of ProvSQL, as we consider they

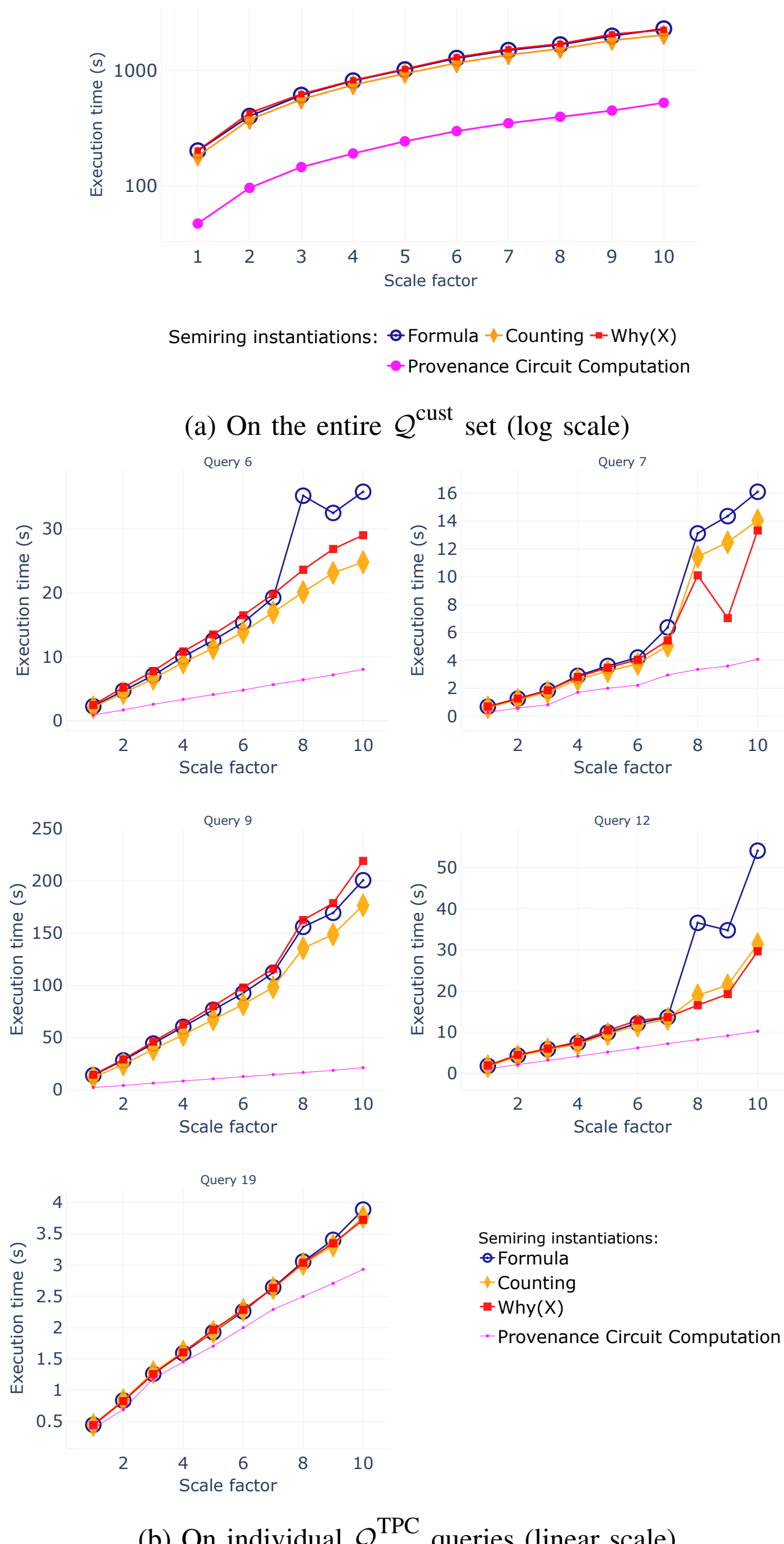

(a) On the entire $\mathcal{Q}^{\text{cust}}$ set (log scale)

(b) On individual $\mathcal{Q}^{\text{TPC}}$ queries (linear scale)

Figure 2: Scalability of semiring instantiations in ProvSQL

are representative to some possible optimizations. Figure 3 (left) shows the scalability of three different versions of ProvSQL, implementing provenance circuits differently. Due to compatibility issues with the older implementations of circuits in ProvSQL we had to to run this experiment on PostgreSQL 14. The current version of ProvSQL (*provsql_mmap*), which uses memory-mapped files, has the fastest running times of all three. Moreover, the running time scales linearly. The shared-memory version of ProvSQL (*provsql_shm*), in which circuits are kept in the PostgreSQL shared memory, does not scale beyond TPC-H scale factor 4. The shared-memory implementation is

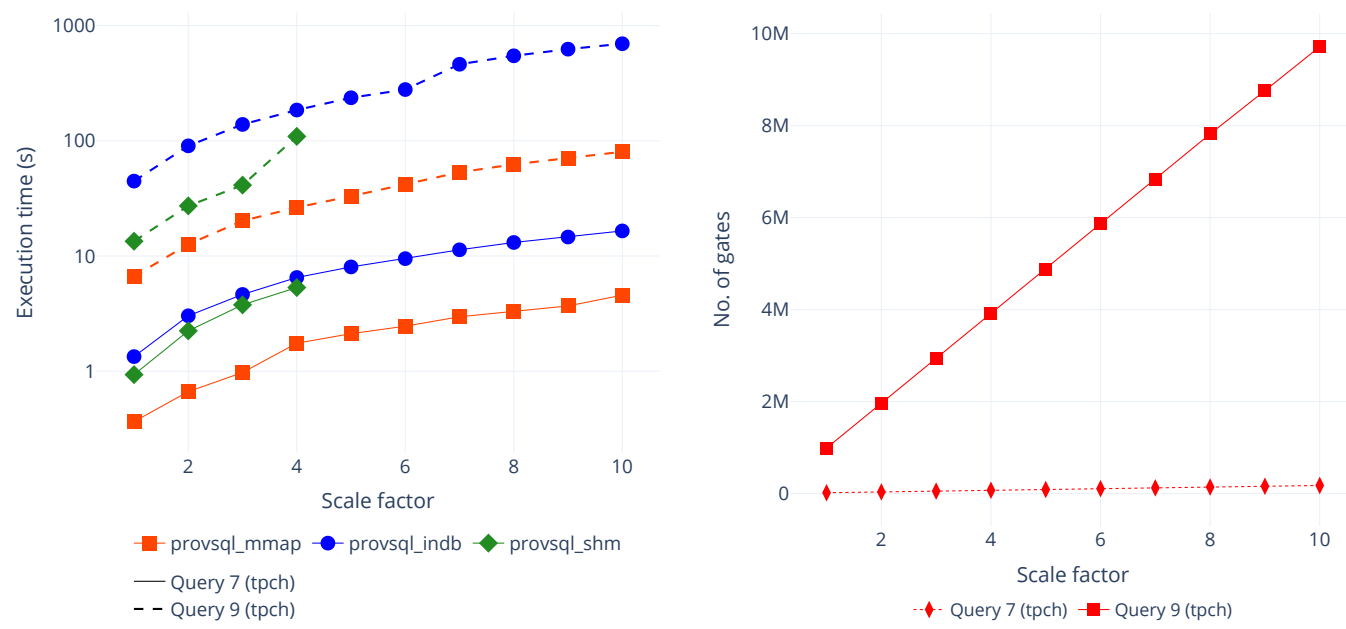

Figure 3: Detailed analysis of ProvSQL on queries 7 and 9 from $\mathcal{Q}^{\text{TPC}}$ for different TPC-H scale factors. Left: scalability and performance of different ProvSQL versions (log scale). Right: number of extra gates of the provenance circuit.

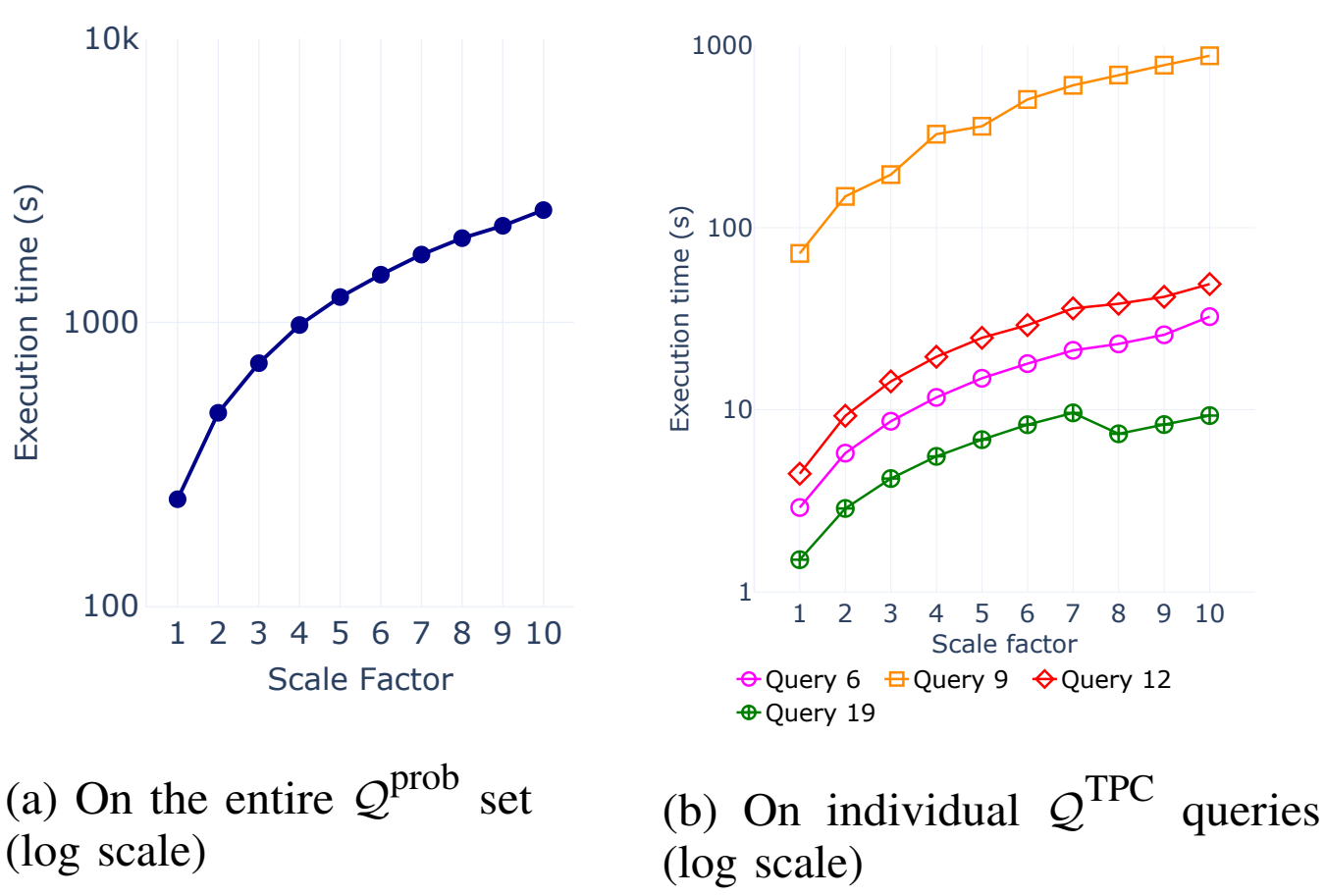

(a) On the entire $\mathcal{Q}^{\text{prob}}$ set (log scale)

(b) On individual $\mathcal{Q}^{\text{TPC}}$ queries (log scale)

Figure 4: Scalability of probability computation in ProvSQL

also slower than the memory-mapped implementation; this is due to the impossibility to parallelize the PL/pgSQL functions that use LOCK on *aggregate values*. Although the in-database implementation of ProvSQL (*provsql_indb*), where provenance circuits are stored as PostgreSQL tables, scales linearly, it has the slowest running times.

Figure 3 (right) shows the number of gates created by ProvSQL in the provenance circuit on evaluation of the query. We can see that the number of gates created at each scale factor is significantly higher when executing query 9 of $\mathcal{Q}^{\text{TPC}}$ than query 7. This is because, even though queries 7 and 9 of $\mathcal{Q}^{\text{TPC}}$ have the same number of joins and aggregations, query 9 is more selective than query 7. This also explains why the execution time for query 9 is significantly higher than query 7, and suggests that reducing the number of gates created in the provenance circuit is a good direction for future optimizations.

*Probability evaluation:* Query 9 in $\mathcal{Q}^{\text{cust}}$ has 8 joins and ProvSQL goes out of memory while computing probabilities for it (more precisely, the knowledge compiler d4 called by ProvSQL runs out of memory). We omit query 9 from $\mathcal{Q}^{\text{cust}}$ and denote this new query set as $\mathcal{Q}^{\text{prob}}$. Figure 4a shows that ProvSQL scales linearly with the dataset size, when computing

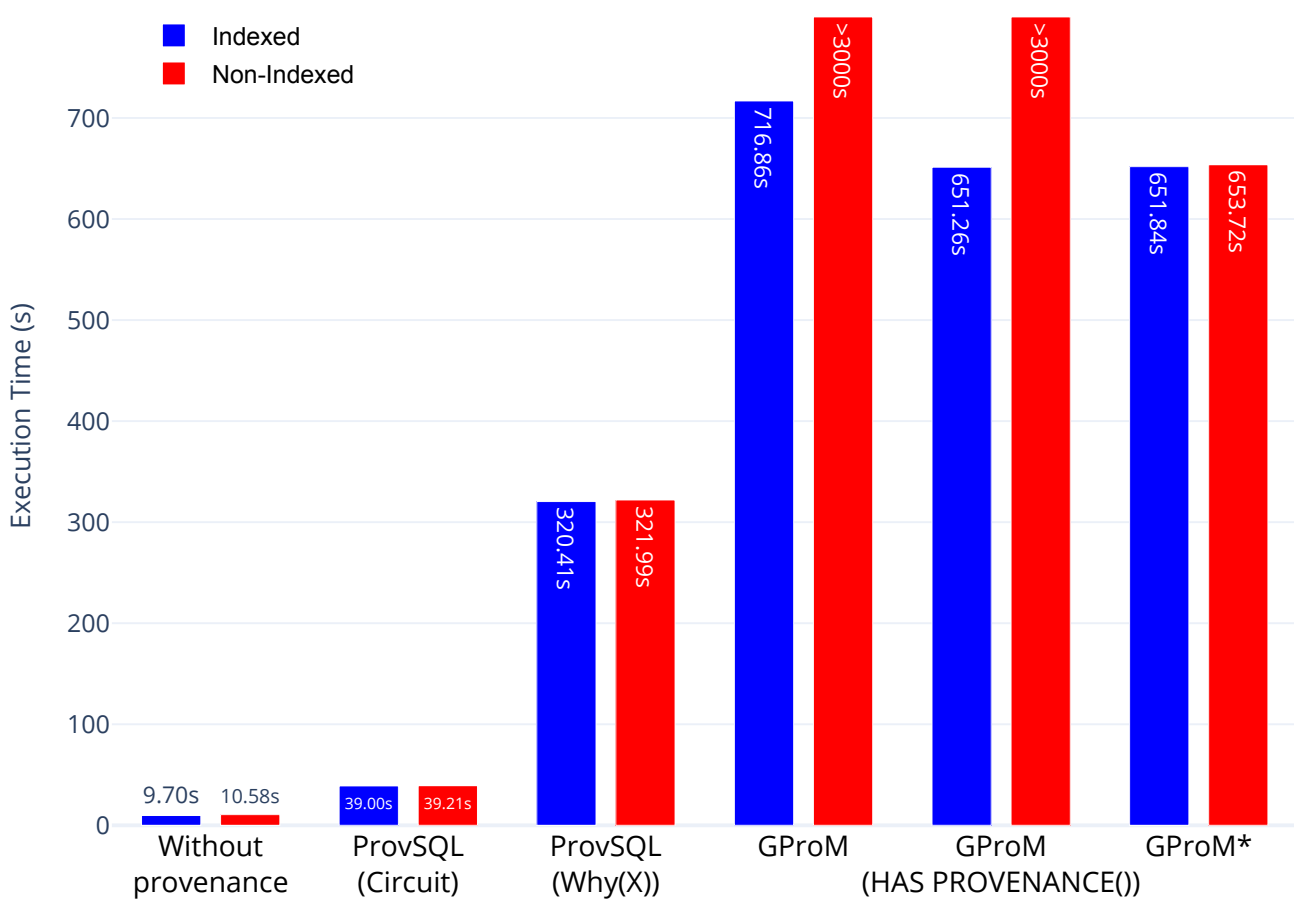

Figure 5: GProM vs ProvSQL on 1 GB TPC-H dataset (with and without indexing) for $\mathcal{Q}^{\text{GProm}}$

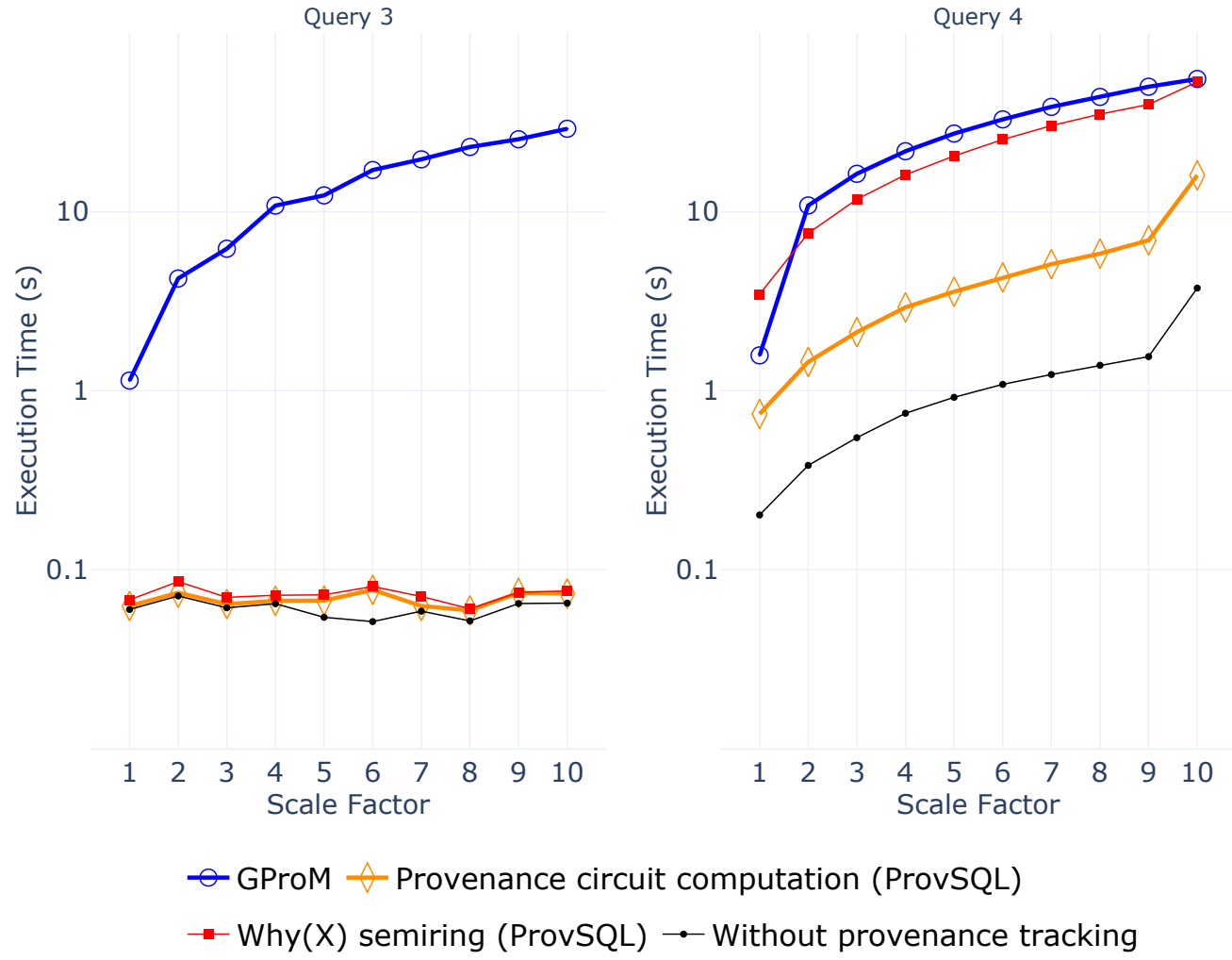

Figure 6: Scalability of GProM vs ProvSQL on query 3 and 4 from $\mathcal{Q}^{\text{TPC*}}$ (log scale)

probabilities on query set $\mathcal{Q}^{\text{prob}}$.

Figure 4b shows generally linear scalability trends across queries 6, 9, 12, 19 from $\mathcal{Q}^{\text{TPC}}$. The exception is query 19: some optimizer driven fluctuations generate counter-intuitive performance *gains* between scale factors 7 and 9. For queries 1 and 7 (not shown here) ProvSQL goes out of memory when computing probabilities, for the same reasons as outlined in the provenance overhead experiment.

### D. GProM vs ProvSQL for Provenance

We now compare the performance of ProvSQL to that of GProM for computing the provenance of query results.

We compare the execution times of provenance tracking for GProM and ProvSQL on $\mathcal{Q}^{\text{GProm}}$, consisting of all queries in $\mathcal{Q}^{\text{cust}}$ that GProM supports (see Table III) in Figure 5. By using GProM's `HAS PROVENANCE()` construct on the primary keys of the tables used in a query, GProM can compute provenance expressions that closely resemble the form of why-provenance defined in semiring-based provenance models. However, in TPC-H, the primary keys are defined separately for each individual table. They cannot be used as provenance annotations because provenance annotations, by construction, should be globally unique across the entire database. Since GProM does not use global identifiers, we created composite primary keys for each table in the TPC-H dataset by prefixing the table name to the primary keys and call `HAS PROVENANCE()` on the composite primary keys.

We first consider the impact of indexing on the performance of these systems. In TPC-H, no indices (primary keys, unique constraints, etc) are defined by default, and it is usually advised not to use indexes to compare the performance of classical relational database systems on this dataset. In small-scale datasets, the benefit of indexing on TPC-H queries is actually negligible. We can observe this in Figure 5 (scale factor 1) for PostgreSQL runtime and provenance computation using ProvSQL. However for GProM, indexing dramatically reduces the execution time. Post-indexing, GProM yields better results when using `HAS PROVENANCE()`. The bars labeled GProM* shows normalized execution times of GProM (with `HAS PROVENANCE()`) rewritings, evaluated directly within PostgreSQL, without relying on query evaluation in GProM. This allows direct comparison of the query rewriting approaches between GproM and ProvSQL. The reported times are the total execution time of both query rewriting and query evaluation.

In Figure 6 we compare the scalability of provenance tracking in GProM with ProvSQL on queries 3 and 4 from $\mathcal{Q}^{\text{TPC*}}$. These queries are simple queries without aggregation. For query 3 the number of output tuples is limited to 20 and provenance tracking is highly efficient across scale factors for ProvSQL. The semiring instantiations incur negligible overhead over baseline execution times and circuit computation times. Even though GProM underperforms relatively to ProvSQL, it scales linearly. For query 4 there are no constraints on the number of output tuples. Both ProvSQL and GProM scale linearly. The semiring instantiations in ProvSQL incur constant overhead throughtout the scale factors. For query 4 GProM's execution times are slightly higher (except at scale factor 1).

### E. MayBMS vs ProvSQL for Probability

We compare the scalability of ProvSQL and MayBMS when it comes to probability computation on uncertain tuple-independent databases, excluding queries involving set operations and aggregations, that MayBMS does not support (see Table III). The queries used in this experiment were rewritten for MayBMS to adhere to the syntax constraints of this system.

MayBMS implements efficient query plans for *safe queries* [12], something ProvSQL does not. We first select 4 queries that are not safe (11, 14, 15 and 17), from our set of custom queries for which MayBMS can compute probabilities; this query set is denoted as $\mathcal{Q}^{\text{MayBMS}}$. Figure 7a shows that both ProvSQL and MayBMS perform similarly on all 4 queries,

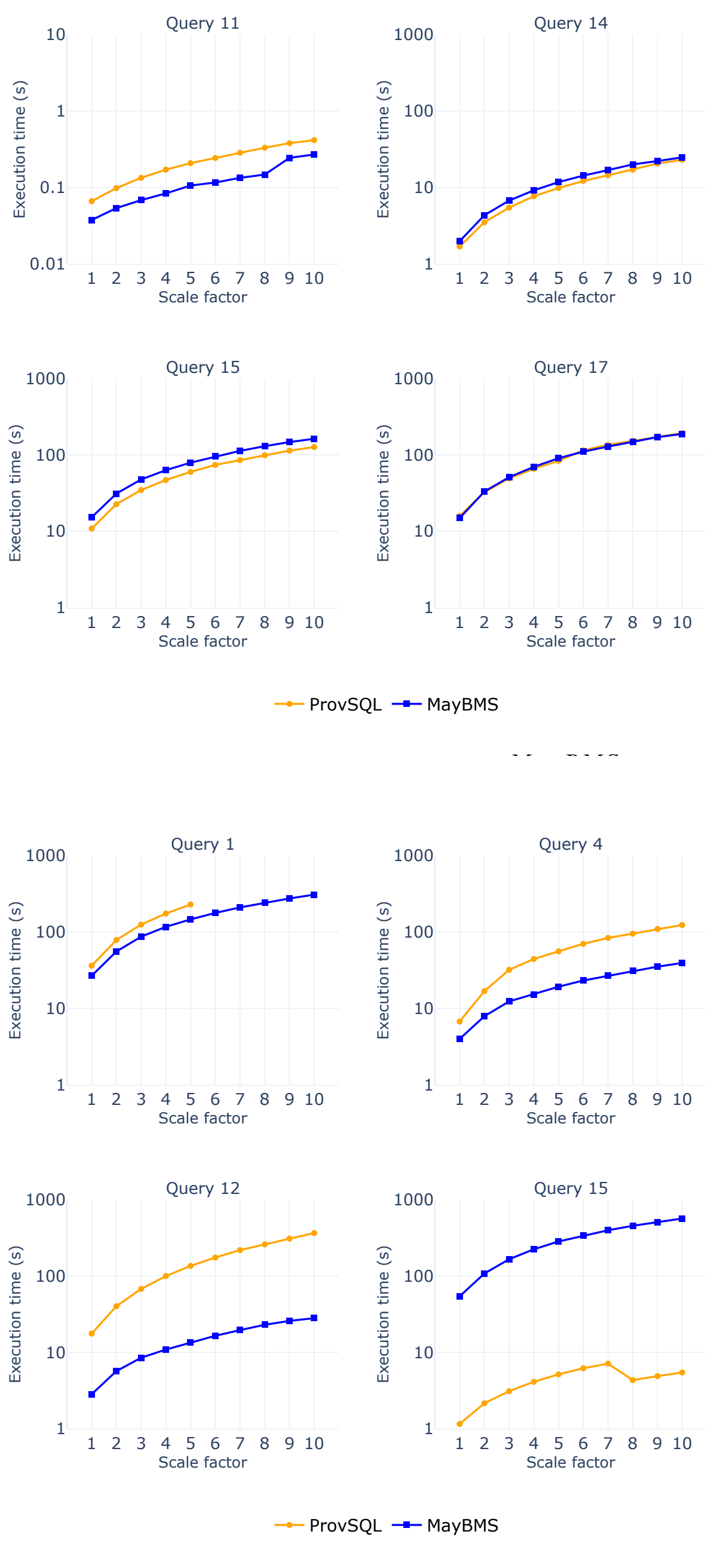

(b) On queries 1, 4, 12 and 15 from $\mathcal{Q}^{\text{TPC}^*}$ (log scale)

Figure 7: Scalability of MayBMS vs ProvSQL

exhibiting linear growth with TPC-H scale factors. ProvSQL performs slightly better on queries 14 and 15; ProvSQL and MayBMS have nearly the same execution times for query 17.

We also compared ProvSQL with MayBMS by carrying out probability computations on the modified TPC-H queries, which were used to benchmark MayBMS. These are the largest subqueries from the original TPC-H queries 1, 4, 12, and 15 but without aggregations and inequality joins. Figure 7b shows how MayBMS and ProvSQL scales against the TPC-H scale factors. For query 1, ProvSQL goes out of memory beyond scale factor 5. Both systems scale linearly with dataset size, but MayBMS performs noticeably better than ProvSQL on queries 1, 4 and 12. This difference in their performance is mainly because MayBMS and ProvSQL have different approaches when it comes to probability evaluation. In this experiment all queries are safe and MayBMS employs safe query plans which often results in efficient probability evaluation. ProvSQL does not implement the safe query approach. It is designed to be a generic system that allows provenance tracking and also offers probability computation by means of Boolean provenance circuits. When provenance circuits exceed manageable limits it resorts to knowledge compilers, which may be unable to exploit the structure of the provenance formula. In some queries, however, such as query 15, this mechanism allows ProvSQL to compute probabilities faster than MayBMS.

## VII. Conclusion

ProvSQL's data model, architecture, and generic design makes it suitable to compute a large range of provenance annotations and to provide a generic solution for query evaluation in probabilistic databases. ProvSQL supports a larger subset of SQL than other existing systems and is actively maintained. Its performance is suitable for multi-GB databases; overhead of provenance computation is often constant (with the notable exception of aggregate queries) and, despite the #P-hardness of the problem, it is often possible to compute probabilities of query outputs in acceptable time.

Many perspectives for improvement exist. First, augmenting the support of SQL is a necessity for general applicability; this includes relatively simple cases such as subqueries in `WHERE` clauses or `OUTER JOIN` queries, nested aggregations, and much more complex cases such as `WITH RECURSIVE` queries which requires changes to the query executor. Second, when evaluating safe queries over tuple-independent databases, the intensional approach to probabilistic query evaluation is suboptimal. It is unknown whether the safe plan algorithm [12] can be turned into an intensional approach, but some results in this direction [54] are a first step toward safe query plans.

## Acknowledgements

This research is part of the program DesCartes and is supported by the National Research Foundation, Prime Minister's Office, Singapore under its Campus for Research Excellence and Technological Enterprise (CREATE) program. It is also partially supported by DataGEMS, funded by the European Union's Horizon Europe Research and Innovation programme, under grant agreement 101188416. We are also grateful to Charles Paperman for suggesting the memory-mapped implementation of the circuit.

## AI-Generated Content Acknowledgement

We used generative AI tools, like ChatGPT, to draft an initial version of some automation scripts for benchmarks and figures; these scripts were subsequently reviewed and fully adjusted by hand. Some of the Lean formal proofs were also aided by the use of generative AI tools. We did not use any AI system to develop any of the algorithms, theorems, human-readable proofs, or any other technical content of the paper; we did not use any AI system for the implementation of ProvSQL itself.

## Appendix

### Material for Section III (Extended Relational Algebra)

We define the semantics $[\![\cdot]\!]^I$ of relational algebra queries on an instance $I$ over $\mathcal{D}$ by induction:

**relation** for any relation label $R$ in the domain of $\mathcal{D}$, $[\![R]\!]^I \stackrel{\text{def}}{=} I(R)$;

**projection** for $k \in \mathbb{N}$, $q \in \mathsf{RA}_k$, $t_1, \dots, t_n$ terms of max-index $\leqslant k$, $[\![\Pi_{t_1,\dots,t_n}(q)]\!]^I \stackrel{\text{def}}{=} \{\!\!\{\, (t_1(u), \dots, t_n(u)) \mid u \in [\![q]\!]^I \,\}\!\!\}$;

**selection** for $k \in \mathbb{N}$, $q \in \mathsf{RA}_k$, and $\varphi$ a Boolean combination of (in)equality comparisons between terms of max-index $\leqslant k$, $[\![\sigma_\varphi(q)]\!]^I \stackrel{\text{def}}{=} \{\!\!\{\, u \mid u \in [\![q]\!]^I, \varphi(u) \,\}\!\!\}$;

**cross product** for $k_1, k_2 \in \mathbb{N}$, $q_1 \in \mathsf{RA}_{k_1}$, $q_2 \in \mathsf{RA}_{k_2}$, $[\![q_1 \times q_2]\!]^I \stackrel{\text{def}}{=} [\![q_1]\!]^I \times [\![q_2]\!]^I$;

**multiset sum** for $k \in \mathbb{N}$, $q_1, q_2 \in \mathsf{RA}_k$, $[\![q_1 \uplus q_2]\!]^I \stackrel{\text{def}}{=} [\![q_1]\!]^I \uplus [\![q_2]\!]^I$;

**duplicate elimination** for $k \in \mathbb{N}$, $q \in \mathsf{RA}_k$, $[\![\varepsilon(q)]\!]^I$ maps $t$ to 1 if $[\![q]\!]^I(t) > 0$ and to 0 otherwise;

**multiset difference** for $k \in \mathbb{N}$, $q_1, q_2 \in \mathsf{RA}_k$, $[\![q_1 - q_2]\!]^I$ maps $t$ to 0 if $[\![q_2]\!]^I(t) > 0$ and to $[\![q_1]\!]^I(t)$ otherwise;

**aggregation** for $k \in \mathbb{N}$, $q \in \mathsf{RA}_k$, distinct $(i_j)_{1 \leqslant j \leqslant k}$, terms $t_1, \dots t_n$ of max-index $\leqslant k$, functions $f_1, \dots, f_n$ from finite multiset of values to values, $[\![\gamma_{i_1,\dots,i_m}[t_1 : f_1, \dots, t_n : f_n](q)]\!]^I$ is:

$$\Big\{ \Big( v_1, \dots, v_m, f_1 \Big( \{\!\!\{\, t_1(u) \mid u \in [\![q]\!]^I, (u_{i_1}, \dots, u_{i_m}) = (v_1, \dots, v_m) \,\}\!\!\} \Big), \dots, f_n \Big( \{\!\!\{\, t_n(u) \mid u \in [\![q]\!]^I, (u_{i_1}, \dots, u_{i_m}) = (v_1, \dots, v_m) \,\}\!\!\} \Big) \Big) \Big| (v_1, \dots, v_m) \in [\![\varepsilon(\Pi_{\#i_1,\dots,\#i_m}(q))]\!]^I \Big\}.$$

We also give SQL translations of the operators defined in this section, assuming appropriate translations of unnamed positional indices to attribute names within terms (when it makes a difference, we use PostgreSQL syntax variants):

relation

```
SELECT * FROM R
```

projection

```
SELECT t1, t2, ..., tn FROM (q)
```

selection

```
SELECT * FROM (q) WHERE φ
```

cross product

```
SELECT * FROM (q1), (q2)
```

multiset sum

```
q1 UNION ALL q2
```

duplicate elimination

```
SELECT DISTINCT * FROM (q)
```

multiset difference

```
SELECT * FROM (q1) AS temp WHERE ROW(temp.*) NOT IN (q2)
```

aggregation If $\#j$ stands for the name of the $j$th attribute returned by the query $q$:

```
SELECT #i1, ..., #im, f1(t1), ..., fn(tn) FROM (q)
GROUP BY 1, ..., m
```

join

```
SELECT * FROM (q1) JOIN (q2) ON φ
```

or simply

```
SELECT * FROM (q1), (q2) WHERE φ
```

set union

```
q1 UNION q2
```

In addition to relations being unlabeled and untyped, our query semantics slightly differs from SQL in the following way:

- As mentioned, our semantics for multiset difference follows `NOT IN`, not `EXCEPT ALL`.
- We do not consider `NULL` values in any way.
- We do not reproduce some quirks of SQL groupless aggregation: in SQL, `SELECT COUNT(x) FROM R` returns 0 on an empty relation of schema `R(x INT)` while `SELECT SUM(x) FROM R` returns `NULL`. In our semantics, $\gamma_{\#1}[\#1 : f](R)$ returns the empty relation, however $f$ is defined, if evaluated on an instance where $R$ is empty. This is more consistent with aggregation under grouping.

## Material for Section IV (Querying Annotated Relations)

**Proposition 6.** L∃∀N *For a set $X$, $(2^{2^X}, \varnothing, \{\varnothing\}, \cup, \Cup, \setminus)$ is an m-semiring.*

*Proof.* For $R, S, T \in 2^{2^X}$:

(i) $S \cup (T \setminus S) = S \cup T = T \cup S = T \cup (S \setminus T)$.

(ii) $R \setminus (S \cup T) = \{x \in R \mid x \notin S, x \notin T\} = \{x \in R \setminus S \mid x \notin T\} = (R \setminus S) \setminus T$.

(iii) $S \setminus S = \varnothing \setminus S = \varnothing$. ☐

**Theorem 10.** L∃∀N *Let $\mathcal{D}$ be a database schema, $q$ any extended relational algebra query over $\mathcal{D}$, $\mathbb{K}$ an appropriate algebraic structure, and $\hat{I}$ a $\mathbb{K}$-instance over $\mathcal{D}$. Let $\hat{q}$ be the query rewritten from $q$ by applying the rewriting rules (R1)–(R5) recursively bottom up. Then $\langle\!\langle q \rangle\!\rangle^{\hat{I}} = \llbracket \hat{q} \rrbracket^{\hat{I}}$.*

*Proof.* We proceed by induction on the structure of $q$.

- **relation** for any relation label $R$ in the domain of $\mathcal{D}$, $\llbracket \hat{R} \rrbracket^{\hat{I}} = \llbracket R \rrbracket^{\hat{I}} = \hat{I}(R) = \langle\!\langle R \rangle\!\rangle^{\hat{I}}$.
- **projection** for $k \in \mathbb{N}$, $q \in \mathsf{RA}_k$, $t_1, \ldots, t_n$ of terms of max-index $\leqslant k$,

$$\llbracket \widehat{\Pi_{t_1,\ldots,t_n}(q)} \rrbracket^{\hat{I}} = \llbracket \Pi_{t_1,\ldots,t_n,\#(k+1)}(\hat{q}) \rrbracket^{\hat{I}} = \{\!\!\{ (t_1(u), \ldots, t_n(u), \alpha) \mid (u, \alpha) \in \llbracket \hat{q} \rrbracket^{\hat{I}} \}\!\!\}$$
$$= \{\!\!\{ (t_1(u), \ldots, t_n(u), \alpha) \mid (u, \alpha) \in \langle\!\langle q \rangle\!\rangle^{\hat{I}} \}\!\!\} = \langle\!\langle \Pi_{t_1,\ldots,t_n}(q) \rangle\!\rangle^{\hat{I}}.$$

- **selection** for $k \in \mathbb{N}$, $q \in \mathsf{RA}_k$, and $\varphi$ a Boolean combination of (in)equality comparisons involving terms of max-index $\leqslant k$,

$$\llbracket \widehat{\sigma_\varphi(q)} \rrbracket^{\hat{I}} = \llbracket \sigma(\varphi)(\hat{q}) \rrbracket^{\hat{I}} = \{\!\!\{ u' \mid u' \in \llbracket \hat{q} \rrbracket^{\hat{I}}, \varphi(u') \}\!\!\} = \{\!\!\{ (u, \alpha) \mid (u, \alpha) \in \langle\!\langle q \rangle\!\rangle^{\hat{I}}, \varphi(u) \}\!\!\} = \langle\!\langle \sigma_\varphi(q) \rangle\!\rangle^{\hat{I}}.$$

  Indeed, since $\varphi$ does not involve "$\#(k+1)$", $\varphi(u') = \varphi((u, \alpha)) = \varphi(u)$.
- **cross product** for $k_1, k_2 \in \mathbb{N}$, $q_1 \in \mathsf{RA}_{k_1}$, $q_2 \in \mathsf{RA}_{k_2}$,

$$\begin{aligned}
\llbracket \widehat{q_1 \times q_2} \rrbracket^{\hat{I}} &= \llbracket \Pi_{\#1,\ldots,\#k_1,\#(k_1+2),\ldots,\#(k_1+k_2+1),\#(k_1+1)\otimes\#(k_1+k_2+2)}(\hat{q}_1 \times \hat{q}_2) \rrbracket^{\hat{I}} \\
&= \{\!\!\{ (u_1, \ldots, u_{k_1}, v_1, \ldots, v_{k_2}, \alpha \otimes \beta) \mid (u_1, \ldots, u_{k_1}, \alpha, v_1, \ldots, v_{k_2}, \beta) \in \llbracket \hat{q}_1 \times \hat{q}_2 \rrbracket^{\hat{I}} \}\!\!\} \\
&= \{\!\!\{ (u_1, \ldots, u_{k_1}, v_1, \ldots, v_{k_2}, \alpha \otimes \beta) \mid (u_1, \ldots, u_{k_1}, \alpha, v_1, \ldots, v_{k_2}, \beta) \in \llbracket \hat{q}_1 \rrbracket^{\hat{I}} \times \llbracket \hat{q}_2 \rrbracket^{\hat{I}} \}\!\!\} \\
&= \{\!\!\{ (u_1, \ldots, u_{k_1}, v_1, \ldots, v_{k_2}, \alpha \otimes \beta) \mid (u_1, \ldots, u_{k_1}, \alpha, v_1, \ldots, v_{k_2}, \beta) \in \langle\!\langle q_1 \rangle\!\rangle^{\hat{I}} \times \langle\!\langle q_2 \rangle\!\rangle^{\hat{I}} \}\!\!\} \\
&= \langle\!\langle q_1 \times q_2 \rangle\!\rangle^{\hat{I}}.
\end{aligned}$$

- **multiset sum** for $k \in \mathbb{N}$, $q_1, q_2 \in \mathsf{RA}_k$,

$$\llbracket \widehat{q_1 \uplus q_2} \rrbracket^{\hat{I}} = \llbracket \hat{q}_1 \uplus \hat{q}_2 \rrbracket^{\hat{I}} = \llbracket \hat{q}_1 \rrbracket^{\hat{I}} \uplus \llbracket \hat{q}_2 \rrbracket^{\hat{I}} = \langle\!\langle q_1 \rangle\!\rangle^{\hat{I}} \uplus \langle\!\langle q_2 \rangle\!\rangle^{\hat{I}} = \langle\!\langle q_1 \uplus q_2 \rangle\!\rangle^{\hat{I}}.$$

- **duplicate elimination** for $k \in \mathbb{N}$, $q \in \mathsf{RA}_k$,

$$\begin{aligned}
\llbracket \widehat{\varepsilon(q)} \rrbracket^{\hat{I}} &= \llbracket \gamma_{1,\ldots,k}[\#(k+1) : \textstyle\bigoplus](\hat{q}) \rrbracket^{\hat{I}} \\
&= \left\{ \left( v_1, \ldots, v_k, \bigoplus \{\!\!\{ \alpha \mid u \in \llbracket \hat{q} \rrbracket^{\hat{I}}, u = (v_1, \ldots, v_k, \alpha) \}\!\!\} \right) \,\middle|\, (v_1, \ldots, v_k) \in \llbracket \varepsilon(\Pi_{1,\ldots,k}(\hat{q})) \rrbracket^{\hat{I}} \right\} \\
&= \left\{ \left( v_1, \ldots, v_k, \bigoplus \{\!\!\{ \alpha \mid u \in \llbracket \hat{q} \rrbracket^{\hat{I}}, u = (v_1, \ldots, v_k, \alpha) \}\!\!\} \right) \,\middle|\, \llbracket \Pi_{1,\ldots,k}(\hat{q}) \rrbracket^{\hat{I}}(v_1, \ldots, v_k) > 0 \right\} \\
&= \left\{ \left( v_1, \ldots, v_k, \bigoplus \{\!\!\{ \alpha \mid u \in \llbracket \hat{q} \rrbracket^{\hat{I}}, u = (v_1, \ldots, v_k, \alpha) \}\!\!\} \right) \,\middle|\, \exists \alpha\, \llbracket \hat{q} \rrbracket^{\hat{I}}(v_1, \ldots, v_k, \alpha) > 0 \right\} \\
&= \left\{ \left( v_1, \ldots, v_k, \bigoplus \{\!\!\{ \alpha \mid u \in \langle\!\langle q \rangle\!\rangle^{\hat{I}}, u = (v_1, \ldots, v_k, \alpha) \}\!\!\} \right) \,\middle|\, \exists \alpha\, \langle\!\langle q \rangle\!\rangle^{\hat{I}}(v_1, \ldots, v_k, \alpha) > 0 \right\} \\
&= \bigcup_{u \mid \exists \alpha (u,\alpha)(u,va) \in \langle\!\langle q \rangle\!\rangle^{\hat{I}}} \left\{ \left( u, \bigoplus_{\alpha \mid (u,\alpha) \in \langle\!\langle q \rangle\!\rangle^{\hat{I}}} \alpha \right) \right\} \\
&= \langle\!\langle \varepsilon(q) \rangle\!\rangle^{\hat{I}}.
\end{aligned}$$

- **multiset difference** for $k \in \mathbb{N}$, $q_1, q_2 \in \mathsf{RA}_k$,

$$
\begin{aligned}
\llbracket \widehat{q_1 - q_2} \rrbracket^{\hat{I}} = \llbracket \quad & \Pi_{\#1,\dots,\#(k+1)}(\hat{q}_1 \bowtie_{\#1=\#(k+1)\wedge\cdots\wedge\#k=\#(2k)} \varepsilon(\Pi_{\#1,\dots,\#k}(\hat{q}_1) - \Pi_{\#1,\dots,\#k}(\hat{q}_2))) \\
& \uplus \Pi_{\#1,\dots,\#k,\#(k+1)\ominus\#(2k+2)}(\hat{q}_1 \bowtie_{\#1=\#(k+2)\wedge\cdots\wedge\#k=\#(2k+1)} \gamma_{\#1,\dots,\#k}[\#(k+1) : \bigoplus](\hat{q}_2))) \\
= \quad & \llbracket \Pi_{\#1,\dots,\#(k+1)}(\hat{q}_1 \bowtie_{\#1=\#(k+1)\wedge\cdots\wedge\#k=\#(2k)} \varepsilon(\Pi_{\#1,\dots,\#k}(\hat{q}_1) - \Pi_{\#1,\dots,\#k}(\hat{q}_2))) \rrbracket^{\hat{I}} \\
& \uplus \llbracket \Pi_{\#1,\dots,\#k,\#(k+1)\ominus\#(2k+2)}(\hat{q}_1 \bowtie_{\#1=\#(k+2)\wedge\cdots\wedge\#k=\#(2k+1)} \gamma_{\#1,\dots,\#k}[\#(k+1) : \bigoplus](\hat{q}_2)) \rrbracket^{\hat{I}} \\
= \quad & \{\!\{ (u,\alpha) \mid (u,\alpha) \in \llbracket \hat{q}_1 \rrbracket^{\hat{I}}, u \in \llbracket \Pi_{\#1,\dots,\#k}(\hat{q}_1) - \Pi_{\#1,\dots,\#k}(\hat{q}_2) \rrbracket^{\hat{I}} \}\!\} \\
& \uplus \{\!\{ (u, \alpha \ominus \beta) \mid (u,\alpha) \in \llbracket \hat{q}_1 \rrbracket^{\hat{I}}, (u,\beta) \in \llbracket \gamma_{\#1,\dots,\#k}[\#(k+1) : \bigoplus](\hat{q}_2)) \rrbracket^{\hat{I}} \}\!\} \\
= \quad & \{\!\{ (u,\alpha) \mid (u,\alpha) \in \llbracket \hat{q}_1 \rrbracket^{\hat{I}}, \nexists\, \beta\, (u,\beta) \in \llbracket \hat{q}_2 \rrbracket^{\hat{I}} \}\!\} \uplus \left\{\!\!\left\{ \left( u, \alpha \ominus \bigoplus_{\beta \mid (u,\beta) \in \llbracket \hat{q}_2 \rrbracket^{\hat{I}}} \beta \right) \mid (u,\alpha) \in \llbracket \hat{q}_1 \rrbracket^{\hat{I}}, \exists \beta (u,\beta) \in \llbracket \hat{q}_2 \rrbracket^{\hat{I}} \right\}\!\!\right\} \\
= \quad & \{\!\{ (u,\alpha) \mid (u,\alpha) \in \langle\!\langle q_1 \rangle\!\rangle^{\hat{I}}, \nexists\, \beta\, (u,\beta) \in \langle\!\langle q_2 \rangle\!\rangle^{\hat{I}} \}\!\} \uplus \left\{\!\!\left\{ \left( u, \alpha \ominus \bigoplus_{\beta \mid (u,\beta) \in \langle\!\langle q_2 \rangle\!\rangle^{\hat{I}}} \beta \right) \middle| (u,\alpha) \in \langle\!\langle q_1 \rangle\!\rangle^{\hat{I}}, \exists \beta (u,\beta) \in \langle\!\langle q_2 \rangle\!\rangle^{\hat{I}} \right\}\!\!\right\} \\
= \quad & \left\{\!\!\left\{ \left( u, \alpha \ominus \bigoplus_{\beta \mid (u,\beta) \in \langle\!\langle q_2 \rangle\!\rangle^{\hat{I}}} \beta \right) \middle| (u,\alpha) \in \langle\!\langle q_1 \rangle\!\rangle^{\hat{I}} \right\}\!\!\right\} = \langle\!\langle q_1 - q_2 \rangle\!\rangle^{\hat{I}}.
\end{aligned}
$$

- **aggregation** for $k \in \mathbb{N}$, $q \in \mathsf{RA}_k$, distinct $(i_j)_{1 \leqslant j \leqslant k}$, terms $t_1, \dots t_n$ of max-index $\leqslant k$, monoid aggregate functions $f_1, \dots, f_n$,

$$
\begin{aligned}
& \llbracket \gamma_{i_1,\dots,i_m}[t_1 : \widehat{f_1, \dots}, t_n : f_n](q) \rrbracket^{\hat{I}} \\
= & \llbracket \gamma_{i_1,\dots,i_m}[t_1 * \#(k+1) : \hat{f}_1, \dots, t_n * \#(k+1) : \hat{f}_n, \#(k+1) : \delta\left(\bigoplus\right)](\hat{q}) \rrbracket^{\hat{I}} \\
= & \left\{ \left( v_1, \dots, v_m, \hat{f}_1\left( \{\!\{ t_1(u)*\alpha \mid (u,\alpha) \in \llbracket \hat{q} \rrbracket^{\hat{I}}, (u_{i_1},\dots,u_{i_m}) = (v_1,\dots,v_m) \}\!\} \right), \dots, \delta\left( \bigoplus_{\substack{(u,\beta) \in \llbracket \hat{q} \rrbracket^{\hat{I}} \\ (u_{i_1},\dots,u_{i_m}) = (v_1,\dots,v_m)}} \beta \right) \right) \middle| (v_1, \dots, v_m) \in \llbracket \varepsilon(\Pi_{\#i_1,\dots,\#i_m}(\hat{q})) \rrbracket^{\hat{I}} \right\} \\
= & \left\{ \left( v_1, \dots, v_m, \hat{f}_1\left( \{\!\{ t_1(u)*\alpha \mid (u,\alpha) \in \langle\!\langle q \rangle\!\rangle^{\hat{I}}, (u_{i_1},\dots,u_{i_m}) = (v_1,\dots,v_m) \}\!\} \right), \dots, \delta\left( \bigoplus_{\substack{(u,\beta) \in \llbracket \hat{q} \rrbracket^{\hat{I}} \\ (u_{i_1},\dots,u_{i_m}) = (v_1,\dots,v_m)}} \beta \right) \right) \middle| (u,\beta) \in \llbracket \hat{q} \rrbracket^{\hat{I}}, (u_{i_1}, \dots, u_{i_m}) = (v_1, \dots, v_m) \right\} \\
= & \left\{ \left( v_1, \dots, v_m, \hat{f}_1\left( \{\!\{ t_1(u)*\alpha \mid (u,\alpha) \in \langle\!\langle q \rangle\!\rangle^{\hat{I}}, (u_{i_1},\dots,u_{i_m}) = (v_1,\dots,v_m) \}\!\} \right), \dots, \delta\left( \bigoplus_{\substack{(u,\beta) \in \langle\!\langle q \rangle\!\rangle^{\hat{I}} \\ (u_{i_1},\dots,u_{i_m}) = (v_1,\dots,v_m)}} \beta \right) \right) \middle| (u,\beta) \in \langle\!\langle q \rangle\!\rangle^{\hat{I}}, (u_{i_1}, \dots, u_{i_m}) = (v_1, \dots, v_m) \right\} \\
= & \left\{ \left( v_1, \dots, v_m, \hat{f}_1\left( \{\!\{ t_1(u)*\alpha \mid (u,\alpha) \in \langle\!\langle q \rangle\!\rangle^{\hat{I}}, (u_{i_1},\dots,u_{i_m}) = (v_1,\dots,v_m) \}\!\} \right), \dots, \delta\left( \bigoplus_{(u,\beta) \in \langle\!\langle \Pi_{\#i_1,\dots,\#i_m}(q) \rangle\!\rangle^{\hat{I}}} \beta \right) \right) \middle| (u,\beta) \in \langle\!\langle \Pi_{\#i_1,\dots,\#i_m}(q) \rangle\!\rangle^{\hat{I}} \right\} \\
= & \left\{ \left( v_1, \dots, v_m, \hat{f}_1\left( \{\!\{ t_1(u)*\alpha \mid (u,\alpha) \in \langle\!\langle q \rangle\!\rangle^{\hat{I}}, (u_{i_1},\dots,u_{i_m}) = (v_1,\dots,v_m) \}\!\} \right), \dots, \delta(\beta) \right) \middle| (u,\beta) \in \langle\!\langle \varepsilon(\Pi_{\#i_1,\dots,\#i_m}(q)) \rangle\!\rangle^{\hat{I}} \right\} \\
= & \langle\!\langle \gamma_{i_1,\dots,i_m}[t_1 : f_1, \dots, t_n : f_n](q) \rangle\!\rangle^{\hat{I}}.
\end{aligned}
$$

□

**Theorem 12.** *For any finite set of variables $X$, probability distribution* Pr *over $X$, $\mathcal{B}[X]$-relation $\hat{I}$ and relational algebra query $q$ without aggregation, for any tuple $t$ with same arity as $q$,* $\Pr(t \in q(\hat{I})) = \Pr\left( \bigvee_{(t,\alpha) \in \langle\!\langle q \rangle\!\rangle^{\hat{I}}} \alpha \right)$.

*Proof.* For a $\mathcal{B}[X]$-relation $\hat{I}$ and a valuation $\nu$ over $X$, we define $\nu(\hat{I})$ to be the (unannotated) relation $J$ obtained from $\hat{I}$ by removing all tuples with annotation $\alpha$ such as $\alpha(\nu) = \bot$ and keeping all other tuples. In other words, $\nu(\hat{I}) = J \iff \Phi_{\hat{I}}(\hat{J})(\nu) = \top$.

For a $\mathcal{B}[X]$-relation $\hat{I}$, a query $q$ without aggregation and a tuple $t$ with same arity as $q$, we define $\mathrm{Prov}(t \in q(\hat{I}))$, the *Boolean provenance of $t$ for $q$ on $\hat{I}$*, to be the Boolean function that maps a valuation $\nu$ to $\top$ if $t \in [\![q]\!]^{\nu(\hat{I})}$ and to $\bot$ otherwise.

We first note that $\Pr(t \in q(\hat{I})) = \Pr(\mathrm{Prov}(t \in q(\hat{I})))$. Indeed:

$$\Pr(\mathrm{Prov}(t \in q(\hat{I}))) = \sum_{\nu | t \in [\![q]\!]^{\nu(\hat{I})}} \Pr(\nu) = \sum_{\hat{J} \subseteq \hat{I}, t \in [\![q]\!]^J} \sum_{\nu | \nu(\hat{I}) = J} \Pr(\nu) = \sum_{\hat{J} \subseteq \hat{I}, t \in [\![q]\!]^J} \sum_{\nu | \nu(\hat{I}) = J} \Pr(\nu)$$

while

$$\Pr(t \in q(\hat{I})) = \sum_{\hat{J} \subseteq \hat{I}, t \in [\![q]\!]^J} \Pr(\hat{J}) = \sum_{\hat{J} \subseteq \hat{I}, t \in [\![q]\!]^J} \Pr[\Phi_{\hat{I}}(\hat{J})] = \sum_{\hat{J} \subseteq \hat{I}, t \in [\![q]\!]^J} \sum_{\nu | \Phi_{\hat{I}}(\hat{J})(\nu) = \top} \Pr(\nu) = \sum_{\hat{J} \subseteq \hat{I}, t \in [\![q]\!]^J} \sum_{\nu | \nu(\hat{I}) = J} \Pr(\nu).$$

It then suffices to show that

$$\mathrm{Prov}(t \in q(\hat{I})) = \bigvee_{(t,\alpha) \in \langle\!\langle q \rangle\!\rangle^{\hat{I}}} \alpha$$

or, in other terms, that

$$t \in [\![q]\!]^{\nu(\hat{I})} \quad \text{if and only if} \quad \exists (t,\alpha) \in \langle\!\langle q \rangle\!\rangle^{\hat{I}}, \alpha(\nu) = \top.$$

We proceed by induction on the structure of $q$.

- **relation** for any relation label $R$ in the domain of $\mathcal{D}$, $t \in [\![R]\!]^{\nu(\hat{I})} \iff t \in \nu(\hat{I})(R) \iff \exists (t,\alpha) \in \hat{I}(R)\, \alpha(\nu) = \top \iff \exists (t,\alpha) \in \langle\!\langle R \rangle\!\rangle^{\hat{I}}, \alpha(\nu) = \top$.
- **projection** for $k \in \mathbb{N}$, $q \in \mathsf{RA}_k$, $t_1, \ldots, t_n$ of terms of max-index $\leqslant k$,

$$\begin{aligned} t \in [\![\Pi_{t_1,\ldots,t_n}(q)]\!]^{\nu(\hat{I})} &\iff \exists u \in [\![q]\!]^{\nu(\hat{I})}, t = (t_1(u), \ldots, t_n(u)) \\ &\iff \exists (u,\alpha) \in \langle\!\langle q \rangle\!\rangle^{\hat{I}}, \alpha(\nu) = \top \wedge t = (t_1(u), \ldots, t_n(u)) \\ &\iff \exists (t,\alpha) \in \langle\!\langle \Pi_{t_1,\ldots,t_n}(q) \rangle\!\rangle^{\hat{I}}, \alpha(\nu) = \top. \end{aligned}$$

- **selection** for $k \in \mathbb{N}$, $q \in \mathsf{RA}_k$, and $\varphi$ a Boolean combination of (in)equality comparisons involving terms of max-index $\leqslant k$,

$$\begin{aligned} t \in [\![\sigma_\varphi(q)]\!]^{\nu(\hat{I})} &\iff t \in [\![q]\!]^{\nu(\hat{I})}, \varphi(u) \\ &\iff \exists (t,\alpha) \in \langle\!\langle q \rangle\!\rangle^{\hat{I}}, \alpha(\nu) = \top \wedge \varphi(u) \\ &\iff \exists (t,\alpha) \in \langle\!\langle \sigma_\varphi(q) \rangle\!\rangle^{\hat{I}}, \alpha(\nu) = \top. \end{aligned}$$

- **cross product** for $k_1, k_2 \in \mathbb{N}$, $q_1 \in \mathsf{RA}_{k_1}$, $q_2 \in \mathsf{RA}_{k_2}$,

$$\begin{aligned} t \in [\![q_1 \times q_2]\!]^{\nu(\hat{I})} &\iff t = (t_1,t_2), t_1 \in [\![q_1]\!]^{\nu(\hat{I})} \wedge t_2 \in [\![q_2]\!]^{\nu(\hat{I})} \\ &\iff t = (t_1,t_2), \exists (t_1,\alpha) \in \langle\!\langle q_1 \rangle\!\rangle^{\hat{I}}, \alpha(\nu) = \top \wedge \exists (t_2,\beta) \in \langle\!\langle q_2 \rangle\!\rangle^{\hat{I}}, \beta(\nu) = \top \\ &\iff t = (t_1,t_2), \exists (t_1,\alpha,t_2,\beta) \in \langle\!\langle q_1 \rangle\!\rangle^{\hat{I}} \times \langle\!\langle q_2 \rangle\!\rangle^{\hat{I}}, (\alpha \hat{\wedge} \beta)(\nu) = \top \\ &\iff \exists (t,\gamma) \in \langle\!\langle q_1 \times q_2 \rangle\!\rangle^{\hat{I}}, \gamma(\nu) = \top. \end{aligned}$$

- **multiset sum** for $k \in \mathbb{N}$, $q_1, q_2 \in \mathsf{RA}_k$,

$$\begin{aligned} t \in [\![q_1 \uplus q_2]\!]^{\nu(\hat{I})} &\iff t \in [\![q_1]\!]^{\nu(\hat{I})} \uplus [\![q_2]\!]^{\nu(\hat{I})} \\ &\iff (\exists (t,\alpha) \in \langle\!\langle q_1 \rangle\!\rangle^{\hat{I}} \alpha(\nu) = \top) \vee (\exists (t,\alpha) \in \langle\!\langle q_2 \rangle\!\rangle^{\hat{I}} \alpha(\nu) = \top) \\ &\iff \exists (t,\alpha) \in \langle\!\langle q_1 \rangle\!\rangle^{\hat{I}} \uplus \langle\!\langle q_2 \rangle\!\rangle^{\hat{I}} \alpha(\nu) = \top \\ &\iff \exists (t,\alpha) \in \langle\!\langle q_1 \uplus q_2 \rangle\!\rangle^{\hat{I}} \alpha(\nu) = \top. \end{aligned}$$

- **duplicate elimination** for $k \in \mathbb{N}$, $q \in \mathsf{RA}_k$,

$$t \in [\![\varepsilon(q)]\!]^{\nu(\hat{I})} \iff t \in [\![q]\!]^{\nu(\hat{I})} \iff \exists (t,\alpha) \in \langle\!\langle q \rangle\!\rangle^{\hat{I}}, \alpha(\nu) = \top \iff \exists (t,\beta) \in \langle\!\langle \varepsilon(q) \rangle\!\rangle^{\hat{I}}, \beta(\nu) = \top$$

  (the last equivalence comes from the fact that a disjunction is true if and only if one of the disjunct is true).

Table IV: Provenance derivation for GProM on $q_{\text{city}}$

| city | prov_personnel_id | prov_personnel_name | prov_personnel_position | prov_personnel_city | prov_personnel_1_id | prov_personnel_1_name | prov_personnel_1_position | prov_personnel_1_city |
|---|---|---|---|---|---|---|---|---|
| Nairobi | 1 | Juma | Director | Nairobi | 2 | Paul | Janitor | Nairobi |
| Paris | 3 | David | Analyst | Paris | 6 | Nancy | HR | Paris |
| Paris | 3 | David | Analyst | Paris | 5 | Aaheli | Double agent | Paris |
| Beijing | 4 | Ellen | Field agent | Beijing | 7 | Jing | Analyst | Beijing |
| Paris | 5 | Aaheli | Double agent | Paris | 6 | Nancy | HR | Paris |

Table V: Provenance derivation for ProvSQL on $q_{\text{city}}$

| city | why | provsql |
|---|---|---|
| Nairobi | {"{Juma,Paul}"} | d1b22232-... |
| Paris | {"{David,Aaheli}","{Aaheli,Nancy}","{David,Nancy}"} | d82a769d-... |
| Beijing | {"{Ellen,Jing}"} | 1a407bcb-... |

- **multiset difference** for $k \in \mathbb{N}$, $q_1, q_2 \in \mathsf{RA}_k$,

$$
\begin{aligned}
t \in \llbracket q_1 - q_2 \rrbracket^{\nu(\hat{I})} &\iff t \in \llbracket q_1 \rrbracket^{\nu(\hat{I})} \wedge t \notin \llbracket q_2 \rrbracket^{\nu(\hat{I})} \\
&\iff (\exists \alpha (t, \alpha) \in \langle\!\langle q_1 \rangle\!\rangle^{\hat{I}}, \alpha(\nu) = \top) \wedge \neg (\exists \beta (t, \beta) \in \langle\!\langle q_2 \rangle\!\rangle^{\hat{I}}, \beta(\nu) = \top) \\
&\iff (\exists \alpha (t, \alpha) \in \langle\!\langle q_1 \rangle\!\rangle^{\hat{I}}, \alpha(\nu) = \top) \wedge \neg \left( \bigvee_{\beta \mid (t,\beta) \in \langle\!\langle q_2 \rangle\!\rangle^{\hat{I}}} \beta(\nu) = \top \right) \\
&\iff (\exists \alpha (t, \alpha) \in \langle\!\langle q_1 \rangle\!\rangle^{\hat{I}}, \alpha(\nu) \wedge \neg \left( \bigvee_{\beta \mid (t,\beta) \in \langle\!\langle q_2 \rangle\!\rangle^{\hat{I}}} \beta(\nu) = \top \right) = \top \\
&\iff \exists (t, \alpha') \in \langle\!\langle q_1 - q_2 \rangle\!\rangle^{\hat{I}}, \alpha'(\nu) = \top. \qquad \square
\end{aligned}
$$

## Provenance and Probability in Different Systems

We explain the query interface for provenance computation in GProM and ProvSQL, and for probabilistic query evaluation in MayBMS and ProvSQL. This helps understanding how these systems represent provenance and uncertain information, and how to produce a benchmark adapted to each system.

We consider again from Example 2 the *Personnel* instance introduced in Table I and the query $q_{\text{city}}$, which can be written in SQL as follows:

```
SELECT DISTINCT p1.city
FROM personnel p1 JOIN personnel p2
ON p1.city=p2.city AND p1.id<p2.id
```

*Provenance computation:* To find the provenance of this query in GProM we need to use the `PROVENANCE OF` construct in the query statement:

```
PROVENANCE OF (SELECT DISTINCT p1.city
  FROM personnel p1 JOIN personnel p2
  ON p1.city=p2.city AND p1.id<p2.id)
```

For this query, GProM produces the provenance derivation in Table IV. GProM attaches multiple additional attributes to the output tuples of a query. In GProM's result, "Paris" appears three times: once due to David and Aaheli, once due to Aaeheli and Nancy, and once due to David and Nancy. Each occurrence of Paris in GProM's output is accompanied by the provenance attributes of the two input tuples that caused it to appear. We see in particular that GProM disregards the `DISTINCT` keyword: the output of the query would be the same without it.

To find similar provenance information for the same query using ProvSQL, we first make the table provenance-aware by using the `add_provenance` UDF and then specialize the generic provenance token, using a provenance mapping *personnel_name* where each tuple is represented by the *name* attribute, to why-provenance:

```
SELECT p1.city, why(provenance(), 'personnel_name')
FROM personnel p1 JOIN personnel p2
ON p1.city=p2.city AND p1.id<p2.id
GROUP BY p1.city
```

Provenance evaluation is seen as a form of aggregation, which explains why we transformed `DISTINCT` into a `GROUP BY`. This results in a more compact representation, as shown in Table V. ProvSQL computes the why-provenance for each output tuple as sets of sets. Of course, in ProvSQL, any (m-)semiring can be used for the computation, not just why-provenance: ProvSQL defines a number of common examples, and a user can add new UDFs defining the operations of the semiring of choice.

Table VI: The *Personnel_prob* table in MayBMS

| id | name | position | city | _v0 | _d0 | _p0 |
|---|---|---|---|---|---|---|
| 1 | Juma | Director | Nairobi | 1008 | 1 | 0.5 |
| 2 | Paul | Janitor | Nairobi | 1009 | 1 | 0.7 |
| 3 | David | Analyst | Paris | 10010 | 1 | 0.3 |
| 4 | Ellen | Field agent | Beijing | 10011 | 1 | 0.2 |
| 5 | Aaheli | Double agent | Paris | 10012 | 1 | 1.0 |
| 6 | Nancy | HR | Paris | 10013 | 1 | 0.8 |
| 7 | Jing | Analyst | Beijing | 10014 | 1 | 0.2 |

Table VII: Result of probability computation in ProvSQL (right) and MayBMS (left)

| city | conf |
|---|---|
| Nairobi | 0.35 |
| Paris | 0.86 |
| Beijing | 0.04 |

| city | prob | provsql |
|---|---|---|
| Nairobi | 0.35 | d1b22232-... |
| Paris | 0.86 | d82a769d-... |
| Beijing | 0.04 | 1a407bcb-... |

*Probability computation:* We now consider the TID relation introduced in Example 14.

To compute probabilities in MayBMS, the following steps have to be followed. We first use **PICK TUPLES** to create a new TID table with probabilities given by some SQL expression. In this new table, named *Personnel_prob*, MayBMS adds three extra attribute: an event variable *_v0*, a decision value *_d0* and a probability *_p0*, as shown in Table VI. In short, each unique event variable translate to a set of mutually exclusive events, independent across event variables; the decision value specifies the event, and the probability is that of the event value. This is an encoding of the compact U-relation model for probabilistic databases [AJKO08], itself very similar to that of probabilistic c-tables [GT06]. MayBMS offers two functions for probabilistic qurery evaluation: `conf()` and `tconf()` for computing the exact probabilities (along with some functions for computing approximations). The difference between these two functions is that `conf()` computes a probability of distinct results (along with a **GROUP BY** clause), while `tconf()` performs no duplicate elimination. We can then use the following query:

```
SELECT city, conf() FROM (
  SELECT p1.city
  FROM personnel_prob p1, personnel_prob p2
  WHERE p1.city=p2.city AND p1.id<p2.id) temp
GROUP BY city
```

whose result in MayBMS is given in Table VII (left), matching what we computed in Example 14. Note that MayBMS does not support the **JOIN** keyword for queries with self-join, so the query needs to be rewritten with a subquery as shown.

In ProvSQL, we first attach a probability on Boolean variables using the `set_prob()` UDF then, at query time, we use the a UDF to perform probabilistic query evaluation:

```
SELECT p1.city, probability_evaluate(provenance())
FROM personnel p1 JOIN personnel p2
ON p1.city=p2.city AND p1.id<p2.id
GROUP BY p1.city
```

The result is shown in Table VII (right) – we notice the result is still a provenance-aware table, with its *provsql* attribute.

## Generating Custom Queries using DSQGEN

The TPC-DS 3.2.0 dataset comes with a query generator, DSQGEN. DSQGEN is more flexible than the query generator from the TPC-H dataset in the sense that the distributions for the scalars in the queries can be written directly in the query template syntax. This allows us to create our own query templates for querying the TPC-H dataset easily. We then use DSQGEN to generate the executable query files from our query templates.

For example let us consider a query to retrieve line items and their corresponding orders for a given ship date range on the TPC-H dataset. We will have to write the query template like below.

```
DEFINE SHIPDATE_MIN = RANDOM(1,121,uniform);
DEFINE SHIPDATE_MAX = RANDOM([SHIPDATE_MIN],121,uniform);
SELECT l.*, o.*
FROM lineitem l, orders o
WHERE l.l_orderkey = o.o_orderkey
```

```
AND l.l_shipdate >= date '1994-12-01'
- interval '[SHIPDATE_MIN]' day
AND l.l_shipdate < date '1995-12-01'
- interval '[SHIPDATE_MAX]' day;
```

We can then use DSQGEN to generate the following executable query file from the query template for a random seed(here, 3983237069).

```
SELECT l.*, o.*
FROM lineitem l, orders o
WHERE l.l_orderkey = o.o_orderkey
AND l.l_shipdate >= date '1994-12-01'
 - interval '98' day
AND l.l_shipdate < date '1995-12-01'
 - interval '117' day;
```

In this way we have written 18 different query templates following the DSQGEN query template syntax.

DSQGEN QUERY TEMPLATES FOR QUERIES IN $\mathcal{Q}^{\text{CUST}}$

```
-- Query 1:
define QUANT=RANDOM(1,50,uniform);
define MODE=TEXT({"REG AIR",1},{"AIR",1},{"RAIL",1},
{"SHIP",1},{"TRUCK",1},{"MAIL",1},{"FOB",1});
define DISC=RANDOM(0,10,uniform);
SELECT l_orderkey, l_partkey, l_suppkey,
l_linenumber, l_linestatus
FROM lineitem
WHERE l_shipmode = '[MODE]'
AND l_quantity > [QUANT]
OR l_discount >= cast([DISC] as float)/100.0;

-- Query 2:
define NATION=RANDOM(0,24,uniform);
define PRICE=RANDOM(10000,100000,uniform);
SELECT DISTINCT o.o_orderdate, o.o_custkey
FROM orders o, customer c
WHERE o.o_custkey = c.c_custkey
AND o.o_totalprice > [PRICE]
AND c.c_nationkey = [NATION]
AND o.o_orderstatus IN ('P','F');

-- Query 3:
define REGION=RANDOM(0,4,uniform);
define COST=RANDOM(1,1000,uniform);
SELECT p.p_name, p.p_mfgr, p.p_partkey, p.p_retailprice
FROM part p
INNER JOIN partsupp ps ON p.p_partkey = ps.ps_partkey
INNER JOIN supplier s  ON ps.ps_suppkey = s.s_suppkey
INNER JOIN nation n    ON s.s_nationkey = n.n_nationkey
INNER JOIN region r    ON n.n_regionkey = r.r_regionkey
WHERE r.r_regionkey = [REGION]
AND ps.ps_supplycost < [COST];

-- Query 4:
define PRICE=RANDOM(10000,100000,uniform);
SELECT DISTINCT c.c_name, c.c_nationkey
FROM customer c, orders o
WHERE c.c_custkey = o.o_custkey
AND o.o_totalprice > [PRICE]
GROUP BY c.c_nationkey,c.c_name;

-- Query 5:
define COST=RANDOM(100,1000,uniform);
define AVAIL=RANDOM(1,8888,uniform);
SELECT s.s_name, s.s_address, s.s_phone
FROM supplier s INNER JOIN partsupp ps
```

```sql
ON s.s_suppkey = ps.ps_suppkey
WHERE ps.ps_supplycost > [COST]
OR (ps.ps_availqty > [AVAIL] AND
ps.ps_supplycost > cast([COST] as float)/2);

-- Query 6:
define A=TEXT({"STANDARD",1},{"SMALL",1},{"MEDIUM",1},
{"LARGE",1},{"ECONOMY",1},{"PROMO",1});
define B=TEXT({"ANODIZED",1},{"BURNISHED",1},{"PLATED",1},
{"POLISHED",1},{"BRUSHED",1});
define C=TEXT({"TIN",1},{"NICKEL",1},{"BRASS",1},
{"STEEL",1},{"COPPER",1});
define SIZE=RANDOM(1,50,uniform);
SELECT part.p_name, part.p_brand
FROM part
WHERE p_type = '[A] [B] [C]'
AND p_size = [SIZE]
GROUP BY part.p_brand, part.p_name;

-- Query 7:
define MODE=TEXT({"REG AIR",1},{"AIR",1},{"RAIL",1},
{"SHIP",1},{"TRUCK",1},{"MAIL",1},{"FOB",1});
define QUANT=RANDOM(1,50,uniform);
SELECT *
FROM lineitem
WHERE l_shipmode = '[MODE]'
AND l_quantity < [QUANT]
AND l_linestatus = 'F';

-- Query 8:
define OP=TEXT({"1-URGENT",1},{"2-HIGH",1},{"3-MEDIUM",1},
{"4-NOT SPECIFIED",1},{"5-LOW",1});
define CLERK=RANDOM(1,9,uniform);
SELECT c.c_name, o.o_orderdate
FROM orders o INNER JOIN customer c ON o_custkey= c_custkey
WHERE o_orderpriority = '[OP]'
AND o_clerk like 'Clerk#00000000[CLERK]%';

-- Query 9:
define NATION=RANDOM(0,24,uniform);
define BAL=RANDOM(1,10000,uniform);
SELECT c.c_name, o.o_orderstatus
FROM customer c, orders o, nation n, region r, part p,
supplier s, partsupp ps, lineitem l
WHERE c.c_custkey = o.o_custkey
AND o.o_orderkey = l.l_orderkey
AND l.l_partkey = ps.ps_partkey
AND ps.ps_suppkey= s.s_suppkey
AND ps.ps_partkey = p.p_partkey
AND s.s_nationkey = n.n_nationkey
AND n.n_regionkey = r.r_regionkey
AND c.c_nationkey = [NATION]
AND c_acctbal > [BAL]
GROUP BY o.o_orderstatus, c.c_name;

-- Query 10:
define NATION=RANDOM(0,24,uniform);
define BAL=RANDOM(1,10000,uniform);
SELECT c.c_name, o.o_orderstatus
FROM customer c, orders o, partsupp ps, lineitem l
WHERE c.c_custkey = o.o_custkey
AND o.o_orderkey = l.l_orderkey
AND l.l_partkey = ps.ps_partkey
AND c.c_nationkey = [NATION]
AND c_acctbal > [BAL]
GROUP BY o.o_orderstatus, c.c_name;
```

```sql
-- Query 11:
define REGION=RANDOM(0,4,uniform);
define BAL=RANDOM(1,10000,uniform);
SELECT s.*
FROM supplier s
INNER JOIN nation n ON  s.s_nationkey = n.n_nationkey
INNER JOIN region r ON  n.n_regionkey = r.r_regionkey
WHERE r.r_regionkey = [REGION]
AND s.s_acctbal < [BAL];

-- Query 12:
define DMIN = RANDOM(1,121,uniform);
define DMAX = RANDOM([DMIN],121,uniform);
define TAX_A = RANDOM(0,3,uniform);
define TAX_B = RANDOM(4,8,uniform);
SELECT l.*, o.o_clerk, o.o_orderdate, o.o_totalprice
FROM lineitem l INNER JOIN orders o
ON l.l_orderkey = o.o_orderkey
WHERE l.l_shipdate >= date '1994-12-01'
- interval '[DMIN]' day
AND l.l_shipdate < date '1996-12-01'
- interval '[DMAX]' day
AND (
l.l_tax <= cast([TAX_A] as float)/100.0
 OR
(l.l_tax > cast([TAX_B] as float)/100.0 AND o.o_orderstatus = 'O')
);

-- Query 13:
define NATION = RANDOM(0,24,uniform);
define P_MIN = RANDOM(10000,100000,uniform);
define P_MAX = RANDOM([P_MIN],200000,uniform);
SELECT c.c_name, o.o_orderstatus, c.c_nationkey, c.c_phone
FROM customer c, orders o
WHERE c.c_custkey = o.o_custkey
AND c.c_nationkey = [NATION]
AND o.o_totalprice >= [P_MIN]
AND o.o_totalprice <= [P_MAX];

-- Query 14:
define REGION=RANDOM(0,4,uniform);
define S_MIN=RANDOM(1,1000,uniform);
define S_MAX=RANDOM([S_MIN],2000,uniform);
SELECT s.s_name, p.p_brand, p.p_name
FROM supplier s, partsupp ps, part p, nation n, region r
WHERE s.s_suppkey = ps.ps_suppkey
AND ps.ps_partkey = p.p_partkey
AND s.s_nationkey = n.n_nationkey
AND n.n_regionkey = r.r_regionkey
AND r.r_regionkey = [REGION]
AND ps.ps_supplycost >= [S_MIN]
AND ps.ps_supplycost <= [S_MAX]
GROUP BY p.p_brand, p.p_name, s.s_name;

-- Query 15:
define P_MIN = RANDOM(10000,100000,uniform);
define P_MAX = RANDOM([P_MIN],200000,uniform);
SELECT DISTINCT c.c_name, o.o_orderkey, l.l_linenumber
FROM customer c, orders o, lineitem l
WHERE c.c_custkey = o.o_custkey
AND o.o_orderkey = l.l_orderkey
AND o.o_totalprice >= [P_MIN]
AND o.o_totalprice <= [P_MAX];

--Query 16:
define SIZE=RANDOM(1,50,uniform);
define M=RANDOM(1,5,uniform);
```

```
define N=RANDOM(1,5,uniform);
SELECT c.c_name AS name, 'Customer' AS type
FROM customer c, orders o, lineitem l, part p
WHERE c.c_custkey = o.o_custkey
AND o.o_orderkey = l.l_orderkey
AND l.l_partkey = p.p_partkey
AND p.p_size = [SIZE]
EXCEPT
SELECT c.c_name AS name, 'Customer' AS type
FROM customer c, orders o, lineitem l, part p
WHERE c.c_custkey = o.o_custkey
AND o.o_orderkey = l.l_orderkey
AND l.l_partkey = p.p_partkey
AND p.p_brand = 'BRAND#[M][N]';

-- Query 17:
define MIN = RANDOM(1,50,uniform);
define MAX = RANDOM([MIN],50,uniform);
SELECT p.p_type, p.p_name, s.s_name, ps.ps_supplycost
FROM part p, partsupp ps, supplier s
WHERE p.p_partkey = ps.ps_partkey
AND ps.ps_suppkey = s.s_suppkey
AND p.p_size >= [MIN]
AND p.p_size <= [MAX]
GROUP BY p.p_type, p.p_name, s.s_name, ps.ps_supplycost ;

-- Query 18:
define NATION = RANDOM(1,25,uniform);
SELECT c.c_name AS name, 'Customer' AS type
FROM customer c
WHERE c.c_nationkey = [NATION]
UNION
SELECT s.s_name AS name, 'Supplier' AS type
FROM supplier s
WHERE s.s_nationkey = [NATION];
```

$Q^{TPC}$

```
--Query 1:
select
l_returnflag,
l_linestatus,
sum(l_quantity) as sum_qty,
sum(l_extendedprice) as sum_base_price,
sum(l_extendedprice * (1 - l_discount)) as sum_disc_price,
sum(l_extendedprice * (1 - l_discount) * (1 + l_tax)) as sum_charge,
avg(l_quantity) as avg_qty,
avg(l_extendedprice) as avg_price,
avg(l_discount) as avg_disc,
count(*) as count_order

from
        lineitem
where
        l_shipdate <= date '1998-12-01' - interval '117' day
group by
        l_returnflag,
        l_linestatus
order by
        l_returnflag,
        l_linestatus;

--Query 6:
select
        sum(l_extendedprice * l_discount) as revenue
from
        lineitem
where
```

```sql
	l_shipdate >= date '1995-01-01'
	and l_shipdate < date '1995-01-01' + interval '1' year
	and l_discount between 0.09 - 0.01 and 0.09 + 0.01
	and l_quantity < 24;

--Query 7:
select
	supp_nation,
	cust_nation,
	l_year,
	sum(volume) as revenue
from
	(
		select
			n1.n_name as supp_nation,
			n2.n_name as cust_nation,
			extract(year from l_shipdate) as l_year,
			l_extendedprice * (1 - l_discount) as volume
		from
			supplier,
			lineitem,
			orders,
			customer,
			nation n1,
			nation n2
		where
			s_suppkey = l_suppkey
			and o_orderkey = l_orderkey
			and c_custkey = o_custkey
			and s_nationkey = n1.n_nationkey
			and c_nationkey = n2.n_nationkey
			and (
				(n1.n_name = 'RUSSIA' and n2.n_name = 'INDIA')
				or (n1.n_name = 'INDIA' and n2.n_name = 'RUSSIA')
			)
			and l_shipdate between date '1995-01-01' and date '1996-12-31'
	) as shipping
group by
	supp_nation,
	cust_nation,
	l_year
order by
	supp_nation,
	cust_nation,
	l_year;

--Query 9:

select
	nation,
	o_year,
	sum(amount) as sum_profit
from
	(
		select
			n_name as nation,
			extract(year from o_orderdate) as o_year,
			l_extendedprice * (1 - l_discount) - ps_supplycost * l_quantity as amount
		from
			part,
			supplier,
			lineitem,
			partsupp,
			orders,
			nation
		where
			s_suppkey = l_suppkey
```

```sql
                                and ps_suppkey = l_suppkey
                                and ps_partkey = l_partkey
                                and p_partkey = l_partkey
                                and o_orderkey = l_orderkey
                                and s_nationkey = n_nationkey
                                and p_name like '%orchid%'
        ) as profit
group by
        nation,
        o_year
order by
        nation,
        o_year desc;

--Query 12:
select
        l_shipmode,
        sum(case
                when o_orderpriority = '1-URGENT'
                        or o_orderpriority = '2-HIGH'
                        then 1
                else 0
        end) as high_line_count,
        sum(case
                when o_orderpriority <> '1-URGENT'
                        and o_orderpriority <> '2-HIGH'
                        then 1
                else 0
        end) as low_line_count
from
        orders,
        lineitem
where
        o_orderkey = l_orderkey
        and l_shipmode in ('TRUCK', 'AIR')
        and l_commitdate < l_receiptdate
        and l_shipdate < l_commitdate
        and l_receiptdate >= date '1997-01-01'
        and l_receiptdate < date '1997-01-01' + interval '1' year
group by
        l_shipmode
order by
        l_shipmode;

--Query 19:
select
        sum(l_extendedprice* (1 - l_discount)) as revenue
from
        lineitem,
        part
where
        (
                p_partkey = l_partkey
                and p_brand = 'Brand#54'
                and p_container in ('SM CASE', 'SM BOX', 'SM PACK', 'SM PKG')
                and l_quantity >= 4 and l_quantity <= 4 + 10
                and p_size between 1 and 5
                and l_shipmode in ('AIR', 'AIR REG')
                and l_shipinstruct = 'DELIVER IN PERSON'
        )
        or
        (
                p_partkey = l_partkey
                and p_brand = 'Brand#51'
                and p_container in ('MED BAG', 'MED BOX', 'MED PKG', 'MED PACK')
                and l_quantity >= 11 and l_quantity <= 11 + 10
```

```
            and p_size between 1 and 10
            and l_shipmode in ('AIR', 'AIR REG')
            and l_shipinstruct = 'DELIVER IN PERSON'
    )
    or
    (
            p_partkey = l_partkey
            and p_brand = 'Brand#21'
            and p_container in ('LG CASE', 'LG BOX', 'LG PACK', 'LG PKG')
            and l_quantity >= 28 and l_quantity <= 28 + 10
            and p_size between 1 and 15
            and l_shipmode in ('AIR', 'AIR REG')
            and l_shipinstruct = 'DELIVER IN PERSON'
    );
```

$$Q^{\text{TPC*}}$$

```
--Query1
SELECT l_returnflag, l_linestatus FROM
lineitem WHERE l_shipdate<=date '1998-09-01' GROUP BY l_returnflag, l_linestatus;

--Query3
 SELECT l_orderkey FROM customer, orders, lineitem
 WHERE c_mktsegment='BUILDING' AND c_custkey=o_custkey AND
 l_orderkey=o_orderkey AND o_orderdate<date '1995-03-15' AND
 l_shipdate>date '1995-05-15' GROUP BY l_orderkey,
 o_orderdate, o_shippriority LIMIT 20;

--Query 4
 SELECT o_orderpriority FROM orders, lineitem
 WHERE o_orderdate>=date '1993-07-01' AND o_orderdate<date '1993-10-01'
 AND l_orderkey=o_orderkey AND
 l_commitdate<l_receiptdate GROUP BY o_orderpriority;

--Query12
 SELECT l_shipmode FROM orders, lineitem WHERE
 o_orderkey=l_orderkey AND ( l_shipmode='MAIL' OR l_shipmode='SHIP' )
 AND l_commitdate<l_receiptdate AND
 l_shipdate<l_commitdate AND l_receiptdate>='1992-01-01'
 AND l_receiptdate<'1999-01-01'
 GROUP BY l_shipmode;

--Query 15
 SELECT s_suppkey, s_name, s_address, s_phone
 FROM supplier, lineitem WHERE s_suppkey=l_suppkey
 AND l_shipdate>=date '1991-10-10' AND
 l_shipdate<date '1992-01-10' GROUP BY s_suppkey, s_name, s_address, s_phone;
```

## References for the Appendix